\documentclass[english,final, twoside]{IEEEtran}
\usepackage[T1]{fontenc}
\usepackage[latin9]{inputenc}
\usepackage{array}
\usepackage{float}
\usepackage{calc}
\usepackage{bm}
\usepackage{multirow}
\usepackage{amsmath}
\usepackage{amssymb}
\usepackage{graphicx}
\usepackage{esint}
\PassOptionsToPackage{normalem}{ulem}
\usepackage{ulem}

\makeatletter

\providecommand{\tabularnewline}{\\}
\floatstyle{ruled}
\newfloat{algorithm}{tbp}{loa}
\providecommand{\algorithmname}{Algorithm}
\floatname{algorithm}{\protect\algorithmname}

  \newtheorem{asm}{Assumption}
  \newenvironment{asmQED}{\begin{asm}}{~\hfill\IEEEQEDclosed\end{asm}}
  \newtheorem{remrk}{Remark}
  \newtheorem{thmQED}{Theorem}
  \newenvironment{lyxThmQED}{\begin{thmQED}}{~\hfill\IEEEQEDclosed\end{thmQED}}
  \newtheorem{lemQED}{Lemma}
  \newenvironment{lyxLemQED}{\begin{lemQED}}{~\hfill\IEEEQEDclosed\end{lemQED}}
  \newtheorem{defQED}{Definition}
  \newenvironment{lyxDefQED}{\begin{defQED}}{~\hfill\IEEEQEDclosed\end{defQED}}

\usepackage{cite}
\usepackage{subfigure}
\usepackage[level=0]{wgroup_message}
\usepackage{flushend}

\author{Junting~Chen,~\IEEEmembership{Student~Member,~IEEE}        and~Vincent~K.~N.~Lau,~\IEEEmembership{Fellow,~IEEE}

\thanks{This paper was accepted in IEEE Journal on Selected Areas in Communications, special issue on 5G wireless communication systems.}

\thanks{The authors are with the Department of Electronic and Computer Engineering (ECE), The Hong Kong University of Science and Technology (HKUST), Hong Kong (e-mail: \{eejtchen, eeknlau\}@ust.hk).}}%

\makeatother

\usepackage{babel}
\begin{document}

\title{Two-Tier Precoding for FDD Multi-cell Massive MIMO Time-Varying Interference
Networks }

\maketitle
\begin{abstract}
Massive MIMO is a promising technology in future wireless communication
networks. However, it raises a lot of implementation challenges, for
example, the huge pilot symbols and feedback overhead, requirement
of real-time global CSI, large number of RF chains needed and high
computational complexity. We consider a two-tier precoding strategy
for multi-cell massive MIMO interference networks, with an outer precoder
for inter-cell/inter-cluster interference cancellation, and an inner
precoder for intra-cell multiplexing. In particular, to combat with
the computational complexity issue for the outer precoding, we propose
a low complexity online iterative algorithm to track the outer precoder
under time-varying channels. We follow an optimization technique and
formulate the problem on the Grassmann manifold. We develop a low
complexity iterative algorithm, which converges to the global optimal
solution under static channels. In time-varying channels, we propose
a compensation technique to offset the variation of the time-varying
optimal solution. We show with our theoretical result that, under
some mild conditions, perfect tracking of the target outer precoder
using the proposed algorithm is possible. Numerical results demonstrate
that the two-tier precoding with the proposed iterative compensation
algorithm can achieve a good performance with a significant complexity
reduction compared with the conventional two-tier precoding techniques
in the literature.\end{abstract}
\begin{keywords}
Massive MIMO, Two-tier Precoding, Tracking Algorithm, Optimization,
Grassmann Manifold

\end{keywords}

\section{Introduction}

\label{sec:intro}

Massive MIMO is a promising technology to meet the future capacity
demand in wireless cellular networks. Equipped with a large number
of antennas, the system has a sufficient number of degrees of freedom
(DoF) to exploit the \emph{spatial multiplexing gains} for intra-cell
users and to mitigate the \emph{inter-cell interference}. However,
the corresponding beamforming (precoder) designs for such multiuser
MIMO (MU-MIMO) interference networks are challenging even in traditional
MIMO systems with a small number of antennas. In \cite{Foschini:2006fk},
the inter-cell interference is mitigated by using coherently coordinated
transmission (CCT) from multiple base stations (BSs) to each user,
using commonly shared global channel state information (CSI). In \cite{Dahrouj:2010kx},
the beamformers are jointly optimized among BSs, where the uplink-downlink
duality is used to obtain the global CSI in a time-division duplex
(TDD) system. Using alternative optimization techniques, WMMSE algorithm
is proposed in \cite{Shi:2011vn} with the objective to maximize the
weighted sum rate for multi-cell systems. Moreover, interference alignment
(IA) approaches were used in \cite{Zhuang:2011ly,Guillaud:2011fk}
for downlink interference cellular networks.

We consider the beamforming design for frequency-division duplex (FDD)%
\footnote{FDD is still a major duplexing technique in the near future, especially
for macro-coverage applications.%
} massive MIMO systems with a large number of antennas $N_{t}$. Unlike
conventional multi-cell MU-MIMO networks, where the schemes in \cite{Foschini:2006fk,Dahrouj:2010kx,Shi:2011vn,Zhuang:2011ly,Guillaud:2011fk}
may be easily implemented, FDD massive MIMO systems induce a lot of
practical issues: (i) \emph{huge pilot symbols and feedback overheads},
(ii) \emph{large number of RF chains}, (iii) \emph{real-time global
CSI sharing,} and (iv) \emph{huge computational complexity} for precoders
at the BSs. For instance, the required number of independent pilot
symbols for transmit side CSI (CSIT) estimation at the mobile scales
as $\mathcal{O}(N_{t})$, and so as the CSIT feedback overheads. In
addition, as $N_{t}$ scales up, the number of RF chains also scales
up, which induces a high fabrication cost and power consumption. Although
the dynamic antenna switching techniques \cite{Sanayei:2004zr,Berenguer:2005ys}
may reduce the required number of RF chains, those solutions did not
fully utilize the benefits of the extra antennas. Moreover, there
is signaling latency over the backhauls and it is highly difficult
to acquire global real-time CSIT for precoding. Finally, the computational
complexity for the precoding algorithms scales quickly with $N_{t}$,
and low complexity precoding algorithms are needed for massive MIMO
systems.

In this paper, we address all the above difficulties by considering
a \emph{two-tier precoding with subspace alignments}. This is motivated
by the clustering behavior of the user terminals. As illustrated in
Fig. \ref{fig:signal-model}, the users in the same cluster may share
the same scattering environment, and hence, they may have similar
spatial channel correlations. Whereas, users from a different cluster
may have different spatial channel correlations. Therefore, we can
decompose the MIMO precoder at the BS into an \emph{outer precoder}
and an \emph{inner precoder}. The outer precoder is used to mitigate
\emph{inter-cell }and\emph{ inter-cluster }interference based on the
statistical channel spatial covariance. Since the spatial correlations
are slowly varying, the outer precoder can be computed on a slower
timescale. On the other hand, the inner precoder is used for spatial
multiplexing of intra-cluster users on the dimension-reduced subspace
spanned by the outer precoder. As a result, the inner precoders are
adaptive to the local real-time CSIT at the BS and can be computed
in a faster timescale. Using the proposed \emph{two-tier precoding
structure}, we shall illustrate in Section \ref{sub:motivation-two-tier}
that the aforementioned technical issues (i)-(iii) associated with
large $N_{t}$ can be substantially alleviated. 

In \cite{Adhikary:2012vn,Adhikary:2013kx}, a zero-forcing based two-tier
precoding has been proposed for single cell massive MIMO systems.
The outer precoders are computed using a block diagonalization (BD)
algorithm. However, it requires a high complexity for computing the
outer precoder, and the tracking issues for the outer precoder under
time-varying channels were not addressed. In fact, the computational
complexity is a serious concern in massive MIMO systems as the number
of antennas scales to very large. For example, in the BD algorithm
proposed in \cite{Adhikary:2012vn,Adhikary:2013kx}, we need to apply
a series of matrix manipulations including SVD to a number of $N_{t}\times N_{t}$
channel covariance matrices each time we update the outer precoder,
and the associated complexity is $\mathcal{O}(N_{t}^{3})$. In addition,
deriving a low complexity iterative algorithm for the BD solution
in \cite{Adhikary:2012vn,Adhikary:2013kx} is far from trivial. To
address the complexity issue, we consider online tracking solutions
to exploit the temporal correlation of the channel matrices. There
is a body of literature for iterative subspace tracking algorithms,
for example, gradient-based algorithms \cite{Comon:1990uq,Yang:1995pi,Utschick:2002mi,Poon:2003zr},
power iteration based algorithms \cite{Hua:1999bh} and the algorithms
based on Krylov subspace approximations \cite{xu1994fast,Tong:2012qo}.
Moreover, the author in \cite{niu2009interference} proposed an iterative
subspace tracking precoder design for MIMO cellular networks. However,
these algorithms have not fully exploited the channel temporal correlations
to enhance the tracking. In this paper, we propose a compensated subspace
tracking algorithm for the online computation of the outer precoder.
The algorithm is derived by solving an optimization problem formulated
on the Grassmann manifold, and its tracking capability is enhanced
by introducing a compensation term that estimates and offsets the
motion of the target signal subspace. Using a control theoretical
approach, we also characterize the tracking performance of the online
outer precoding algorithm in time-varying massive MIMO systems. We
show that, under mild technical conditions, perfect tracking (with
zero convergence error) of the target outer precoder using the proposed
compensation algorithm is possible, despite the channel covariance
matrix being time-varying. In general, we demonstrate with numerical
results that the proposed two-tier precoding algorithm has a good
system performance with low signaling overhead and low complexity
of $\mathcal{O}(N_{t}^{2})$.

The rest of the paper is organized as follows. Section \ref{sec:system-model}
introduces the massive MIMO channel model and the signal model. Section
\ref{sec:two-tier-precoding} illustrates the two-tier precoding techniques.
Section \ref{sec:iterative-algorithm} derives the iterative algorithm
for tracking the outer precoder, where the associated convergence
analysis is given in Section \ref{sec:convergence-analysis}. Numerical
results are given in Section \ref{sec:numerical} and Section \ref{sec:conclustion}
gives the concluding remarks.

\emph{Notations}: We use lower case bold font to denote vectors and
upper case bold font for matrices. $\mathbf{I}_{N}$ denotes the $N\times N$
identity matrix. For matrices $\mathbf{A}\in\mathbb{C}^{N\times p}$
and $\mathbf{B}\in\mathbb{C}^{N\times q}$, $\left[\mathbf{A}\,\mathbf{B}\right]$
denotes a $N\times(p+q)$ concatenated matrix, whose first $p$ columns
are given by $\mathbf{A}$ and the last $q$ columns are given by
$\mathbf{B}$.

\section{System Model}

\label{sec:system-model}

\subsection{Massive MIMO Channel Model with Local Spatial Scattering}

\begin{figure}
\begin{centering}
\includegraphics[width=0.75\columnwidth]{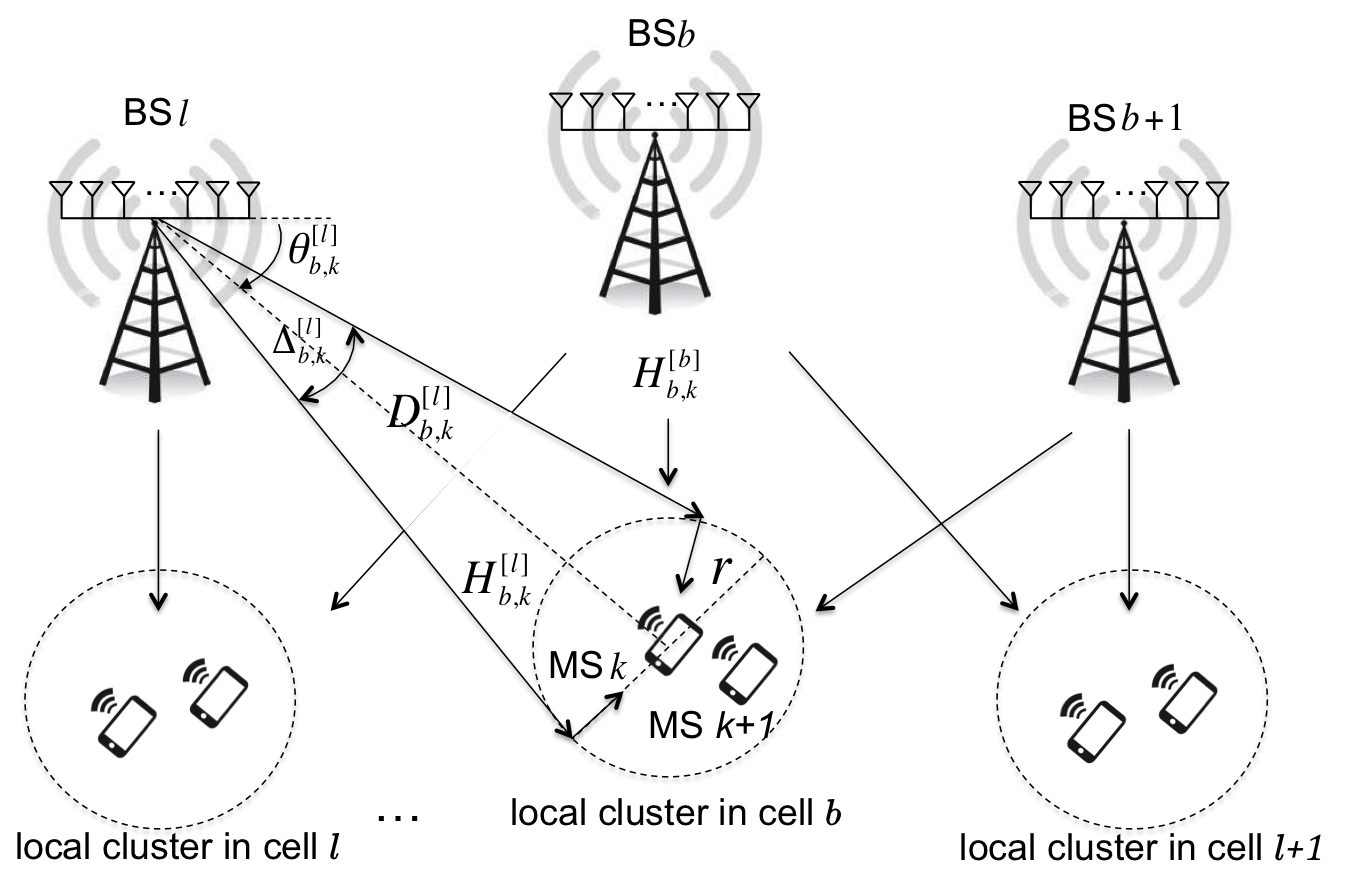}
\par\end{centering}

\caption{\label{fig:signal-model}A multiuser MIMO cellular network. The channel
is modeled by a one-ring local scattering model, where the MS is surrounded
by a scattering ring with radius $r$.}
\end{figure}

We consider a cellular network with $G$ BSs, and the $b$-th ($1\leq b\leq G$)
BS serves $K_{b}$ MSs. The MSs are clustered together, and without
lost of generality, we assume each BS serves one cluster of MSs%
\footnote{Note that the extension to the case of multiple clusters is very straight
forward. Hence we only focus on the single cluster case to simplify
the notation.%
}. Each BS has $N_{t}$ antennas and each user has $N_{r}$ antennas.
The downlink channel from the $l$-th BS to the $k$-th MS in the
$b$-th cell is given by $\mathbf{H}_{b,k}^{[l]}\in\mathbb{C}^{N_{r}\times N_{t}}$.
The receive signal at the MS $k$ in cell $b$ is given by 
\[
\mathbf{y}_{b,k}=\mathbf{H}_{b,k}^{[b]}\mathbf{x}^{[b]}+\sum_{l=1,l\neq b}^{G}\mathbf{H}_{b,k}^{[l]}\mathbf{x}^{[l]}+\mathbf{n}_{b,k}
\]
where $\mathbf{x}^{[l]}\in\mathbb{C}^{d_{l}}$ is the symbol transmitted
at the BS $l$, $d_{l}$ is the number of data streams transmitted
by BS $l$, and $\mathbf{n}_{b,k}\sim\mathcal{CN}(0,\mathbf{I}_{N_{r}})$
is the additive complex Gaussian noise.

In the massive MIMO system, where the BS has a large number of antennas
($N_{t}\gg1$) and is placed on the top of a building, there is usually
not enough local scattering surrounding the BS. Correspondingly, it
has also been shown by channel measurements that most of the signal
energy is localized over the azimuth direction \cite{Correia:2001nx}.
Therefore, we consider the one-ring local scattering model \cite{Abdi:2002ys,Zhang:2007zr}
to characterize the massive MIMO channel. As illustrated in Fig. \ref{fig:signal-model},
the local scattering surrounding the MS is modeled by a ring with
radius $r$; whereas, the transmit signal from the BS shapes a narrow
\emph{angular spread} (AS) denoted as $\triangle_{b,k}^{[l]}\approx2\tan^{-1}(r/D_{b,k}^{[l]})$,
where $D_{b,k}^{[l]}$ is the distance between BS $l$ and MS $k$
in cell $b$. Let $\theta_{b,k}^{[l]}$ be the angle of departure
(AoD) of a path from BS $l$ to MS $k$ in cell $b$. We use the von-Mises
model to characterize the \emph{power azimuth spectrum} (PAS) w.r.t.
$\theta_{b,k}^{[l]}$ \cite{Abdi:2002oq,Abdi:2002ys} as follows:
\begin{equation}
\mathbb{P}_{\theta,(b,k)}^{[l]}(\theta_{b,k}^{[l]})=\frac{\exp\left[\kappa_{b,k}^{[l]}\cos(\theta_{b,k}^{[l]}-\overline{\theta}_{b,k}^{[l]})\right]}{2\pi J_{0}(\kappa_{b,k}^{[l]})},\label{eq:PAS}
\end{equation}
where $\overline{\theta}_{b,k}^{[l]}$ is the mean angle of the AoD,
$J_{0}$ is the zero-th order modified Bessel function and $\kappa_{b,k}^{[l]}=(2\triangle_{b,k}^{[l]})^{-2}$
characterizes the AS at BS $l$ in the direction of MS $k$ in cell
$b$. \mysubnote{R1-A3.c} Denote $\mathbf{T}_{b,k}^{[l]}$ as the
corresponding transmit spatial correlation matrix at BS $l$. The
$(p,q)$-th entry of the matrix $\mathbf{T}_{b,k}^{[l]}$, which describes
the spatial correlations between the $p$-th and $q$-th antenna elements
at BS $l$ \cite{Forenza:2007cr}, is defined as: 
\begin{equation}
\left[\mathbf{T}_{b,k}^{[l]}\right]_{(p,q)}=\int_{-\pi}^{\pi}e^{j\left[\phi_{p}^{[l]}(\theta)-\phi_{q}^{[l]}(\theta)\right]}\mathbb{P}_{\theta,(b,k)}^{[l]}(\theta)d\theta\label{eq:chann-spatial-correlation}
\end{equation}
where $\phi_{p}^{[l]}(\theta)-\phi_{q}^{[l]}(\theta)$ accounts for
the phase difference between the $p$-th and $q$-th antenna elements
over the azimuth direction $\theta$ at BS $l$. 


We assume that MSs within the same cluster have the same channel statistical
parameters $\overline{\theta}_{b,k}^{[l]}$ and $\kappa_{b,k}^{[l]}$,
i.e., $\overline{\theta}_{b,k}^{[l]}=\overline{\theta}_{b,j}^{[l]}$
and $\kappa_{b,k}^{[l]}=\kappa_{b,j}^{[l]}$, $\forall j$. As a result,
the transmit correlation matrices satisfy $\mathbf{T}_{b,k}^{[l]}=\mathbf{T}_{b,j}^{[l]}\triangleq\mathbf{T}_{b}^{[l]}$
for all MS in the scattering cluster of cell $b$. We adopt the following
two-timescale, clustered, and spatial correlated massive MIMO channel
model.
\begin{asmQED}
\label{asm:two-timescale-channel-model}\emph{(Two-timescale, Clustered
and Spatial Correlated Channel Model)} The time-varying massive MIMO
channel $\mathbf{H}_{b,k}^{[l]}(j)$ on each subframe $j$ is given
by 
\begin{equation}
\mathbf{H}_{b,k}^{[l]}(j)=\mathbf{H}_{k}^{\omega}(j)\mathbf{T}_{b}^{[l]}(j)^{1/2}\label{eq:two-timescale-channel-model}
\end{equation}
where $\mathbf{H}_{k}^{\omega}(j)$ and $\mathbf{T}_{b}^{[l]}(j)$
are changing in different timescales:

\begin{itemize}

\item \textbf{Small Timescale:} $\mathbf{H}_{k}^{\omega}(j)$ are
identical and independently distributed (i.i.d.) over MSs $k$ and
is time-varying over subframes $j$. Each element of the matrix $\mathbf{H}_{k}^{\omega}$
follows an independent complex Gaussian distribution with zero mean
and unit variance.

\item \textbf{Large Timescale:} The spatial correlation matrix $\mathbf{T}_{b}^{[l]}(j)$
is constant within each super-frame $(n-1)T_{s}<j\leq nT_{s}$, but
changes between consecutive super-frames (i.e., a block of $T_{s}$
subframes). 

\end{itemize} 
\end{asmQED}

\subsection{Signal Model, Interference Mitigation and Challenges}

Denote the precoding matrix for user $k$ in cell $b$ as $\mathbf{V}_{k}^{[b]}\in\mathbb{C}^{N_{t}\times d_{b,k}}$
and the associated receiver shaping matrix as $\mathbf{U}_{b,k}\in\mathbb{C}^{N_{r}\times d_{b,k}}$,
where $d_{b,k}$ is the number of the data streams. Applying the receiver
shaping matrix $\mathbf{U}_{b,k}$ to the signal $\mathbf{y}_{b,k}$
at MS $k$ in cell $b$, the received signal is given by 
\begin{eqnarray}
\hat{\mathbf{y}}_{b,k} & = & \mathbf{U}_{b,k}^{\dagger}\mathbf{H}_{b,k}^{[b]}\mathbf{V}_{k}^{[b]}\mathbf{s}_{k}^{[b]}+\mathbf{U}_{b,k}^{\dagger}\underbrace{\mathbf{H}_{b,k}^{[b]}\sum_{j=1,j\neq k}^{K_{b}}\mathbf{V}_{j}^{[b]}\mathbf{s}_{j}^{[b]}}_{\mbox{\scriptsize intra-cell interference}}\nonumber \\
 &  & \qquad+\mathbf{U}_{b,k}^{\dagger}\underbrace{\sum_{l=1,l\neq b}^{G}\mathbf{H}_{b,k}^{[l]}\sum_{j=1}^{K_{l}}\mathbf{V}_{j}^{[l]}\mathbf{s}_{j}^{[l]}}_{\mbox{\scriptsize inter-cell interference}}+\mathbf{\hat{n}}_{b,k}\label{eq:signal-model}
\end{eqnarray}
where $\hat{\mathbf{n}}_{b,k}=\mathbf{U}_{b,k}^{\dagger}\mathbf{n}_{b,k}$
is still a standard complex Gaussian noise and $\mathbf{s}_{j}^{[l]}\in\mathbb{C}^{d_{l,j}}$
is the data symbol intended for user $j$ in cell $l$. The per BS
power budget is $\sum_{j=1}^{K_{b}}\mbox{tr}\left(\mathbf{V}_{j}^{[b]}\mathbf{V}_{j}^{[b]\dagger}\right)\leq P$.

In the conventional approach, the inter-cell interference mitigation
and intra-cell spatial multiplexing are achieved by a joint design
of the precoders $\mathbf{V}_{k}^{[b]}$ and receiver shaping matrices
$\mathbf{U}_{b,k}$ among all the BSs using, for example, ZF techniques
\cite{Yoo:2006fk,Kim:2012dq}, WMMSE \cite{Shi:2011vn}, IA \cite{Gomadam:2008uq,Guillaud:2011fk},
etc. However, these approaches cannot be directly applied in FDD massive
MIMO cellular systems because:
\begin{itemize}
\item A large number of RF chains ($N_{t}$) are required to perform RF-baseband
translation as well as Analog-to-Digital (A/D) conversion. As a result,
there is a huge cost in hardware design and power consumption. 
\item A huge amount of pilot symbols should be used to estimate the massive
MIMO channels $\mathbf{H}_{b,k}^{[l]}$ (a large matrix), and a huge
CSI feedback overhead is involved.
\item Global real-time CSIT is required for computing $\mathbf{V}_{k}^{[b]}$.
However, the cross link information $\mathbf{H}_{b,k}^{[l]}$ can
only be obtained via message passing among the backhauls connecting
the BS. This induces a huge burden on \mysubnote{R1-A2} the backhaul
and increases the signaling latency.
\end{itemize}

\mysubnote{R1-A3.a}
\begin{remrk}
[Inter-cell Interference in Massive MIMO] It is reported that the
inter-cell interference (ICI) of multi-cell massive MIMO systems can
be asymptotically ignored \cite{marzetta2010noncooperative} using
simple per-cell zero-forcing. One key assumption is that the direct
links and interference links are spatially uncorrelated. However,
such uncorrelation may not hold under local scattering (such as the
one-ring scattering model considered in this paper), and hence, ICI
coordination may be needed for massive MIMO. 

\mysubnote{Can we call it limited scattering? Line-of-sight propagation also satisfies the uncorrelated assumption.}
\end{remrk}

To deal with these challenges, we propose a two-tier precoding in
the next section.

\section{Two-Tier Precoding: Joint Signal and Interference Subspace Alignment }

\label{sec:two-tier-precoding}

In this section, we propose a two-tier precoding structure by exploiting
the limited local scattering and the clustering structure of mobile
users in cellular systems.

\subsection{Two-tier Precoding with Subspace Alignment}

The precoder at BS $b$ to MS $k$ have the two-tier structure given
by: 
\begin{equation}
\mathbf{V}_{k}^{[b]}=\bm{\Phi}^{[b]}\mathbf{F}_{k}^{[b]}\label{eq:two-tier-precoding-structure}
\end{equation}
where $\bm{\Phi}^{[b]}\in\mathbb{C}^{N_{t}\times m_{b}}$ is the outer
subspace precoder that adapts to the large timescale spatial correlations
to mitigate the \emph{inter-cell interference, }and $\mathbf{F}_{k}^{[b]}\in\mathbb{C}^{m_{b}\times d_{b,k}}$
is the inner precoder that utilizes the real-time local CSIT to mitigate
the \emph{intra-cell interference}. The outer precoder $\bm{\Phi}^{[b]}$
is computed in a long-timescale once every super-frame, and the inner
precoder $\mathbf{F}_{k}^{[b]}$ (and the corresponding receiver shaping
matrices $\mathbf{U}_{b,k}$) is computed in a short-timescale once
every subframe. The parameter $m_{b}$ determines the dimension of
the subspace for intra-cell spatial multiplexing. In the massive MIMO
scenario, we have $m_{b}\ll N_{t}$.  

Specifically, the \emph{two-tier precoding with subspace alignment}
is described below:

\begin{itemize}
\item \textbf{Long-Timescale Processing}: In each super-frame, the subspace
precoders $\{\bm{\Phi}^{[b]}\}_{b=1}^{G}$ are chosen as the solution
to the following optimization problem 
\begin{eqnarray}
\min_{\{\bm{\Phi}^{[b]}\}} & \underbrace{\sum_{l,b,k,l\neq b}\mathbb{E}\left\Vert \mathbf{H}_{b,k}^{[l]}\bm{\Phi}^{[l]}\right\Vert _{F}^{2}}_{\mbox{\scriptsize inter-cell interference}}-w\underbrace{\sum_{b,k}\mathbb{E}\left\Vert \mathbf{H}_{b,k}^{[b]}\bm{\Phi}^{[b]}\right\Vert _{F}^{2}}_{\mbox{\scriptsize intra-cell signal energy}}\label{eq:algorithm-outer-precoder}
\end{eqnarray}
subject to $\bm{\Phi}^{[b]\dagger}\bm{\Phi}^{[b]}=\mathbf{I}_{m_{b}}$,
where the expectations are conditioned on the spatial correlation
matrices $\{\mathbf{T}_{b,k}^{[l]}\}$ and $w>0$ is a weight parameter.
\end{itemize}

\mysubnote{R1-A3.d}
\begin{remrk}
Minimizing the first term in (\ref{eq:algorithm-outer-precoder})
only corresponds to the conventional ZF solution. Whereas, minimizing
the second term alone corresponds to the match filter (MF) solution.
The formulation (\ref{eq:algorithm-outer-precoder}) is to strike
a balance between the inter-cell interference leakage and the intra-cell
signal energy in a system with a large but finite number of antennas.
On one hand, for large number of transmit antennas, the second term
dominates and the solution approaches the MF solution. On the other,
for limited number of antennas (such as traditional traditional MIMO),
the first term is significant and the solution approaches the coordinated
ZF solution. The weight $w$ is to adjust the balance between the
inter-cell interference and the direct link signal.
\end{remrk}
\begin{itemize}
\item \textbf{Short-Timescale Processing}: In each subframe, a ZF precoding
\cite{Yoo:2006fk} is used.

\uline{Step 1}: Choose the receiver shaping matrix $\mathbf{U}_{b,k}\in\mathbb{C}^{N_{r}\times d_{b,k}}$
for MS $k$ in cell $b$ by solving 
\begin{equation}
\min_{\{\mathbf{U}_{b,k}\}}\underbrace{\sum_{l\neq b}\left\Vert \mathbf{U}_{b,k}^{\dagger}\mathbf{H}_{b,k}^{[l]}\bm{\Phi}^{[l]}\right\Vert _{F}^{2}}_{\mbox{\scriptsize remaining inter-cell interference}}-w\underbrace{\left\Vert \mathbf{U}_{b,k}^{\dagger}\mathbf{H}_{b,k}^{[b]}\bm{\Phi}^{[b]}\right\Vert _{F}^{2}}_{\mbox{\scriptsize direct link signal}}\label{eq:algorithm-decoder}
\end{equation}
subject to $\mathbf{U}_{b,k}^{\dagger}\mathbf{U}_{b,k}=\mathbf{I}_{d_{b,k}}$,
and feedback the equivalent channel $\hat{\mathbf{H}}_{b,k}^{[b]}=\mathbf{U}_{b,k}^{\dagger}\mathbf{H}_{b,k}^{[b]}\bm{\Phi}^{[b]}$
to BS $b$.

\uline{Step 2}: Concatenate the rows of $\hat{\mathbf{H}}_{b,k}^{[b]}$
for each MS $k$ to form a $(\sum_{k}d_{b,k})\times m_{b}$ matrix
$\widetilde{\mathbf{H}}^{[b]}\triangleq[\hat{\mathbf{H}}_{b,1}^{[b]\dagger}\,\hat{\mathbf{H}}_{b,2}^{[b]\dagger}\,\dots\,\hat{\mathbf{H}}_{b,K_{b}}^{[b]\dagger}]^{\dagger}$.
The inner precoder $\mathbf{F}^{[b]}$ is given by%
\footnote{We assume that the number of data streams $d_{b,k}$ assigned to each
user always satisfy $\sum_{k}d_{b,k}\leq m_{b}$. Hence, with probability
1, the matrix $\widetilde{\mathbf{H}}^{[b]}$ has full row rank.%
} 
\begin{equation}
\mathbf{F}^{[b]}=\sqrt{\frac{P}{d_{b}}}\widetilde{\mathbf{H}}^{[b]\ddagger}=\sqrt{\frac{P}{d_{b}}}\widetilde{\mathbf{H}}^{[b]\dagger}(\widetilde{\mathbf{H}}^{[b]}\widetilde{\mathbf{H}}^{[b]\dagger})^{-1}\label{eq:algorithm-inner-precoder}
\end{equation}
where $\widetilde{\mathbf{H}}^{[b]\ddagger}$ denotes the \emph{pseudo-inverse}
for $\widetilde{\mathbf{H}}^{[b]}$. The inner precoder $\mathbf{F}_{k}^{[b]}$
for MS $k$ is given by the $(\sum_{j=1}^{k-1}d_{b,j}+1)$-th to the
$(\sum_{j=1}^{k}d_{b,j})$-th columns of $\mathbf{F}^{[b]}$.

\end{itemize}
\begin{remrk}
Similar to the outer precoding, the problem (\ref{eq:algorithm-decoder})
tries to strike a balance between the (remaining) inter-cell interference
and the direct link signal. The inner precoder (\ref{eq:algorithm-inner-precoder})
corresponds to the ZF solution in a single cell multiuser MIMO system
\cite{Yoo:2006fk}.
\end{remrk}
%


\subsection{Motivation of the Two-Tier MIMO Precoding and the Complexity Issue}

\label{sub:motivation-two-tier}

\begin{figure}
\begin{centering}
\includegraphics[width=0.8\columnwidth]{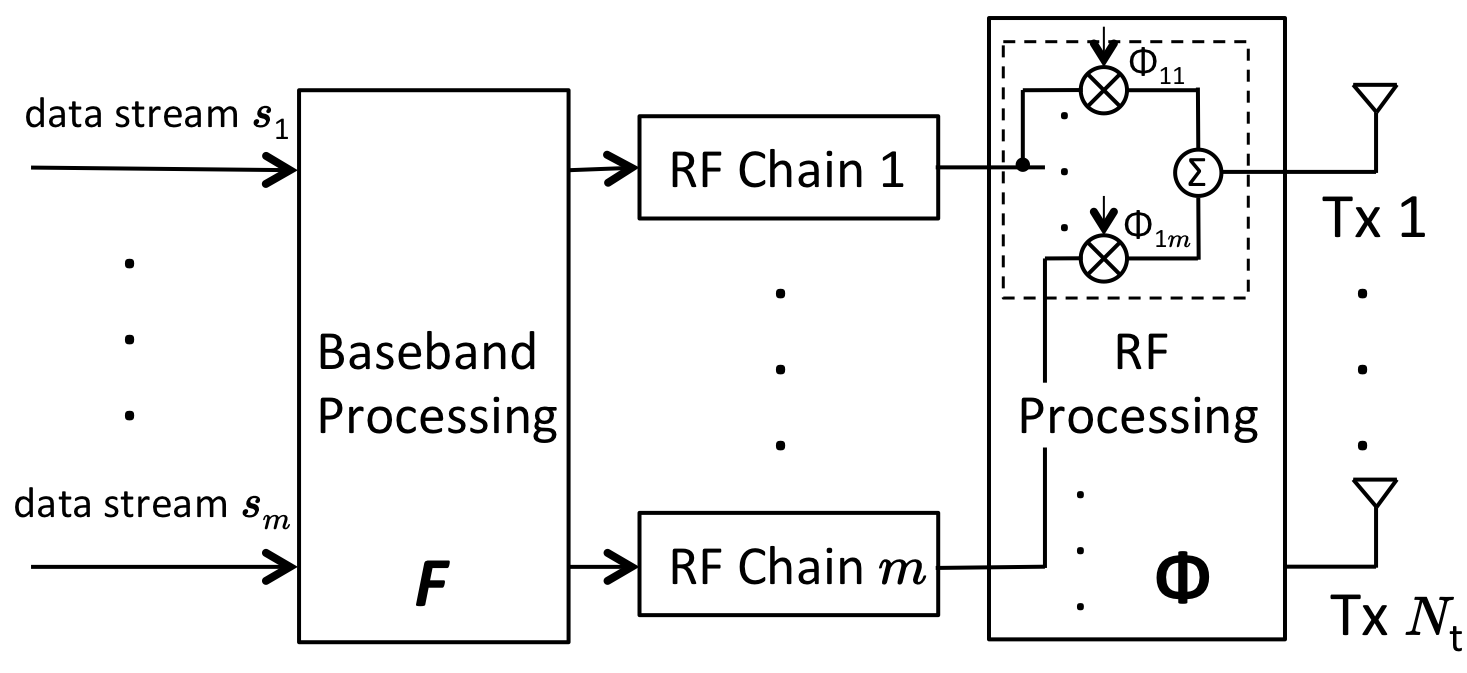}
\par\end{centering}

\caption{\label{fig:RF-chain} An implementation diagram of the two-tier precoding
processing, where only $m$ ($m\ll N_{t}$) RF chains are used.}
\end{figure}

The two-tier MIMO precoding has the following advantages:
\begin{enumerate}
\item \emph{A light demand on pilot symbols and CSI feedback overheads}:
With the outer precoding, the MS only needs to estimate the $N_{r}\times m_{b}$
\emph{effective channel} $\mathbf{H}_{b,k}^{[b]}\bm{\Phi}^{[b]}$
and feedback the $d_{b,k}\times m_{b}$ channel matrix $\mathbf{U}_{b,k}^{\dagger}\mathbf{H}_{b,k}^{[b]}\bm{\Phi}^{[b]}$.
Instead of directly working on the $N_{r}\times N_{t}$ channel matrix
$\mathbf{H}_{b,k}^{[b]}$, there is a huge saving on the pilot symbols
for channel estimation and the CSI feedback loading.
\item \emph{A relatively small number of RF chains required}: With the limited
local scattering around the BS, there is only a few active eigen-modes
for the massive MIMO channel, and a small number of spatial multiplexing
data streams $\sum_{k}d_{b,k}$ can be supported. Therefore, we do
not need to implement $N_{t}$ RF chains. Instead, only $m_{b}$ RF
chains are required, where $\sum_{k}d_{b,k}<m_{b}\ll N_{t}$ and the
outer precoder $\bm{\Phi}^{[b]}$ can be implemented using the RF
phase shifting network \cite{zhang2005variable,sudarshan2006channel}
as illustrated in Fig. \ref{fig:RF-chain}. 
\item \emph{Only statistical global CSI required}: The inner precoder $\mathbf{F}^{[b]}$
only requires the local CSI between the BS and its serving MSs. To
update the outer precoder $\bm{\Phi}^{[b]}$ in the long-timescale,
only the knowledge of channel statistics $\mathbb{E}\left[\mathbf{H}_{l,k}^{[b]^{\dagger}}\mathbf{H}_{l,k}^{[b]}\right]$
is required. As a result, the performance is insensitive to backhaul
latency among the BSs. 
\end{enumerate}

Having addressed the practical issues (i)-(iii) raised in Section
\ref{sec:intro}, we now focus on the computational complexity issue
in (iv). Note that, with the dimension reduction for the inner precoder,
the computation for the outer precoder dominates the complexity. We
first investigate the solution property for the outer precoding problem
(\ref{eq:algorithm-outer-precoder}).
\begin{lyxThmQED}
[Solution to the Outer Precoder]\label{thm:solution-outer-precoder}
The optimal solution $\bm{\Phi}_{*}^{[b]}\in\mathbb{C}^{N_{t}\times m_{b}}$
to the outer precoding problem (\ref{eq:algorithm-outer-precoder})
is given by the eigenvectors corresponding to the $m_{b}$ smallest
eigenvalues of the covariance matrix 
\begin{align}
\mathbf{Q}^{[b]} & \triangleq\sum_{l\neq b}\sum_{k=1}^{K_{l}}\mathbb{E}\left(\mathbf{H}_{l,k}^{[b]\dagger}\mathbf{H}_{l,k}^{[b]}\right)-w\sum_{k=1}^{K_{b}}\mathbb{E}\left(\mathbf{H}_{b,k}^{[b]\dagger}\mathbf{H}_{b,k}^{[b]}\right)\label{eq:interference-signal-covariance-BS}
\end{align}
for each BS $b$.
\end{lyxThmQED}
\begin{proof}
Please refer to Appendix \ref{app:pf-thm-the-outer-precoder-solution}
for the proof.
\end{proof}

Although Theorem \ref{thm:solution-outer-precoder} gives a closed
form expression for $\bm{\Phi}^{[b]}$, computing $\bm{\Phi}^{[b]}$
still require a huge computation complexity of $\mathcal{O}(N_{t}^{3})$.
For example, using SVD for the $N_{t}\times N_{t}$ covariance matrix
$\mathbf{Q}^{[b]}$ requires $\mathcal{O}(N_{t}^{3})$ arithmetic
operations. We will address the computation complexity issue in Section
\ref{sec:iterative-algorithm} and \ref{sec:convergence-analysis}.

\subsection{Achievable Per-cell DoF of Two-Tier Precoding in Massive MIMO}

\label{sub:achievable}

In this section, we characterize the performance of the two-tier precoding
by evaluating its achievable DoF per-cell. The DoF can be interpreted
as the number of data streams or the asymptotic throughput performance
that can be supported in the massive MIMO systems at high SNR \cite{Gomadam:2008uq,Suh:2008tg}.
Denote $C_{b}(P;\{\mathbf{H}_{b,k}^{[l]}\})$ as the sum throughput
of cell $b$. The per-cell DoF of the massive MIMO system is defined
as $\Gamma=\frac{1}{G}\sum_{b=1}^{G}\lim_{P\to\infty}\frac{C_{b}(P;\{\mathbf{H}_{b,k}^{[l]}\})}{\log P}$.

For simplicity, we consider a symmetric massive MIMO network, where
each cell has the same number of MSs, $K_{b}=K$, the same rank $\Upsilon_{b}=\Upsilon<N_{t}$
of transmit spatial correlation matrices $\mathbf{T}_{b}^{[l]}$,
and $m_{b}=\min\{\Upsilon,KN_{r}\}$ for all $b$. We derive the network
DoF of the two-tier precoding in the symmetric massive MIMO system
below. 

\begin{lyxThmQED}
\label{thm:Achievable-DoF} \emph{(Per-cell DoF of Symmetric Massive
MIMO Systems with Two-tier Precoding)} For a symmetric massive MIMO
network $(N_{t},N_{r,}K)^{G}$, where all the transmit correlation
matrices have rank $\Upsilon$, if $N_{t}\geq G\max\{\Upsilon,\min\{\Upsilon,KN_{r}\}\}$,
then 

\begin{itemize}

\item the per-cell DoF of the proposed two-tier precoding is given
by $\Gamma_{\mbox{\scriptsize two}}=\min\{\Upsilon,KN_{r}\},$

\item the per-cell DoF of conventional one-tier interference alignment
(with global real-time CSIT) is given by $\Gamma_{\mbox{\scriptsize one}}=\min\{\Upsilon,KN_{r}\}.$

\end{itemize}
\end{lyxThmQED}
\begin{proof}
Please refer to Appendix \ref{app:pf-lem-achievable-DoF} for the
proof.
\end{proof}

As a result, there is no loss of DoF performance using the proposed
two-tier precoding design in a symmetric multi-cell massive MIMO network%
\footnote{Although the DoF result only focuses on a special network topology
region specified by $N_{t}\geq G\max\{\Upsilon,\min\{\Upsilon,KN_{r}\}\}$,
the region does cover the interested application scenario of massive
MIMO systems, since $N_{t}$ is usually very large and $\Upsilon$
is relatively small in massive MIMO systems.%
}. 


\section{Iterative Algorithms for Outer Precoder under Time-Varying Channels}

\label{sec:iterative-algorithm}

To reduce the complexity of finding the global optimal solution for
the outer precoder problem in (\ref{eq:algorithm-outer-precoder}),
one approach is to leverage on the slowly varying nature of the spatial
correlation $\mathbf{T}_{b}^{[l]}[n]$ and to compute the outer precoder
iteratively at every super-frame. A common technique for such iterative
outer precoder is to apply the gradient descent algorithm to solve
problem (\ref{eq:algorithm-outer-precoder}). However, such a ``naive''
method may have a poor convergence performance, because problem (\ref{eq:algorithm-outer-precoder})
is non-convex due to the quadratic equality constraints $\bm{\Phi}^{[b]\dagger}\bm{\Phi}^{[b]}=\mathbf{I}_{m_{b}}$.
Furthermore, problem (\ref{eq:algorithm-outer-precoder}) suffers
from uncountably many non-unique and non-isolated local optima. For
example, if $\bm{\Phi}_{*}^{[b]}$ is one local optimum, then $\bm{\Phi}_{*}^{[b]}\mathbf{M}$
gives another local optimum, where $\mathbf{M}$ is any unitary matrix.
Such a non-isolated property makes it hard to develop iterative algorithms
with fast convergence to the global optimal solution under time varying
channels. Hence, we need to tackle the following challenge, \\
\framebox{\begin{minipage}[t]{1\columnwidth}%
\textbf{Challenge 1}: To derive a low complexity iterative algorithm
which can be shown to converge to the desired solution for the outer
precoder $\bm{\Phi}^{[b]}$ under time-varying channels.%
\end{minipage}}\medskip{}

To deal with the above challenge, we focus on deriving algorithms
on the \emph{Grassmann manifold}, where all the outer precoders $\bm{\Phi}^{[b]}$
that span the same subspace are considered to be equivalent and are
represented by a single point on the Grassmann manifold. As a result,
the local optimum becomes isolated.

\subsection{Transformation of Problem (\ref{eq:algorithm-outer-precoder}) on
Grassmann Manifold}

\label{sub:formulation-grassmann-manifold}

A \emph{Grassmann manifold} $\mbox{Grass}(m,N_{t})$ is the set of
all $m$-dimensional subspaces of $\mathbb{C}^{N_{t}\times N_{t}}$:
$\mbox{Grass}(m,N_{t})=\{\mbox{span}(\bm{\Phi}):\bm{\Phi}\in\mathbb{C}^{N_{t}\times m},\mbox{rank}(\bm{\Phi})=m\}$,
where $\mbox{span}(\bm{\Phi})$ denotes the space spanned by the columns
of the matrix $\bm{\Phi}$. The \emph{Grassmann manifold} $\mbox{Grass}(m,N_{t})$
can be considered as a \emph{topology} embedded on the Euclidean space
$\mathbb{C}^{N_{t}\times m}$ with a mapping $\pi:\mathbb{C}^{N_{t}\times m}\mapsto\mbox{Grass}(m,N_{t})$
that maps each point from the Euclidean space $\mathbb{C}^{N_{t}\times m}$
to the manifold $\mbox{Grass}(m,N_{t})$. For example, all the matrices
$\bm{\Phi}\mathbf{M}\in\mathbb{C}^{N_{t}\times m}$ ($\mathbf{M}\in\mathbb{C}^{m\times m}$
with full rank $m$), which span the same subspace as $\bm{\Phi}$
does, are all mapped to the same element in $\mbox{Grass}(m,N_{t})$
under the mapping $\pi$, i.e., $\pi(\bm{\Phi})=\pi(\bm{\Phi}\mathbf{M})\in\mbox{Grass}(m,N_{t})$.
On the other hand, the inverse mapping $\pi^{-1}(\bm{\Phi})$ represents
the set of matrices in the Euclidean space $\mathbb{C}^{N_{t}\times m}$
that span the same subspace. 

Consider $\widetilde{\bm{\Phi}}=(\bm{\Phi}^{[1]},\dots,\bm{\Phi}^{[G]})$
as an element on the Grassmann manifold, i.e., $\widetilde{\bm{\Phi}}\in\prod_{b}\mbox{Grass}(m_{b},N_{t})$.
The outer precoding problem (\ref{eq:algorithm-outer-precoder}) (see
equation (\ref{eq:algorithm-outer-precoder-reformulation}) in Appendix
\ref{app:pf-thm-the-outer-precoder-solution}) can be reformulated
as \mysubnote{R2-A3} an optimization over the Grassmann manifold
\cite{manton2002optimization,Absil:2009vn} 
\begin{eqnarray}
\min_{\widetilde{\bm{\Phi}}}\;\;\mathcal{I}(\widetilde{\bm{\Phi}})\triangleq\sum_{b=1}^{G} & \mbox{tr}\left[\left(\bm{\Phi}^{[b]\dagger}\bm{\Phi}^{[b]}\right)^{-1}\bm{\Phi}^{[b]\dagger}\mathbf{Q}^{[b]}\bm{\Phi}^{[b]}\right]\label{prob:objective-Grassmann}
\end{eqnarray}
where the \emph{global optimal solution} is defined to be the subspace
$\widetilde{\bm{\Phi}}_{*}\in\prod_{b}\mbox{Grass}(m_{b},N_{t})$
that yields the minimum objective value of (\ref{prob:objective-Grassmann}). 

Note that, there is a substantial difference between the formulation
(\ref{prob:objective-Grassmann}) in the Grassmann manifold and (\ref{eq:algorithm-outer-precoder})
in the Euclidean space. While the objective in (\ref{eq:algorithm-outer-precoder})
is to find a matrix in the Euclidean space that minimizes the interference
and signal utility function $\mathcal{I}$, the problem (\ref{prob:objective-Grassmann})
focuses on choosing the right subspace $\widetilde{\bm{\Phi}}_{*}$,
\mysubnote{R2-A4} which is a unique solution to minimizing $\mathcal{I}$
\footnote{Problem (\ref{prob:objective-Grassmann}) has a unique solution, if,
for $b=1,\dots,G$, the $m_{b}$-th eigenvalue of the covariance matrix
$\mathbf{Q}^{[b]}$ has multiplicity 1. %
}. With this insight, we can derive algorithms to obtain the subspace
precoders $\bm{\Phi}_{*}^{[b]}$ more efficiently.

\subsection{Outer Precoder Tracking Algorithm with Compensations}

\label{sub:algorithm-gradient}

An intuitive way yo derive an algorithm that solves the problem (\ref{prob:objective-Grassmann}),
is to generalize the gradient descent algorithm to the Grassmann manifold:
\begin{equation}
\widetilde{\bm{\Phi}}[n+1]=\widetilde{\bm{\Phi}}[n]+\gamma_{n}F(\widetilde{\bm{\Phi}}[n];\mathcal{Q}[n])\label{eq:gradient-algorithm}
\end{equation}
where $\gamma_{n}$ is the step size, and 
\begin{equation}
F(\widetilde{\bm{\Phi}};\mathcal{Q})\triangleq\nabla\mathcal{I}(\widetilde{\bm{\Phi}};\mathcal{Q})\label{eq:gradient-mapping-F}
\end{equation}
is the gradient iteration mapping on the Grassmann manifold and $\mathcal{Q}[n]=(\mathbf{Q}^{[1]}[n],\,\dots,\mathbf{Q}^{[G]}[n])$
is a collection of the covariance matrices at iteration $n$. The
notation $\mathcal{I}(\widetilde{\bm{\Phi}};\mathcal{Q})$ emphasizes
that $\mathcal{Q}$ is the key parameter that determines the optimal
solution $\widetilde{\bm{\Phi}}_{*}(\mathcal{Q})$.

Although the gradient descent algorithm usually finds a stationary
point under static parameters, it may not be the case under the time-varying
parameter $\mathcal{Q}[n]$. Under time-varying channels, the channel
covariance matrices $\{\mathbf{Q}^{[b]}[n]\}$ are varying in a similar
timescale as the gradient iterations and there is always a convergence
gap between the gradient iterate $\widetilde{\bm{\Phi}}[n]$ and the
time-varying optimal solution $\widetilde{\bm{\Phi}}_{*}(\mathcal{Q}[n])$.
Taking the precoder $\bm{\Phi}^{[b]}$ for BS $b$ as an example:
When the gradient iteration $\bm{\Phi}^{[b]}[n+1]$ gets closer to
the previous target $\bm{\Phi}_{*}^{[b]}[n]$ at the $(n+1)$-th iteration,
the optimal target has already moved to a new position $\bm{\Phi}_{*}^{[b]}[n+1]$,
which contributes to an additional tracking error. 

Intuitively, one way to enhance the tracking of the outer precoder
under time-varying channels is to estimate the motion of the moving
target $\widetilde{\bm{\Phi}}_{*}[n]$ and compensate for it: 
\begin{equation}
\widetilde{\bm{\Phi}}[n+1]=\widetilde{\bm{\Phi}}[n]+\gamma_{n}F(\widetilde{\bm{\Phi}}[n];\mathcal{Q}[n])+\triangle\widehat{\bm{\tilde{\Phi}}_{*}[n]}\label{eq:compensation-algorithm}
\end{equation}
where the compensation term $\triangle\widehat{\bm{\tilde{\Phi}}_{*}[n]}$
is an estimation of the difference of $\widetilde{\bm{\Phi}}_{*}[n+1]-\widetilde{\bm{\Phi}}_{*}[n]$. 

In the following, we illustrate how to derive the gradient mapping
$F(\centerdot)$ and the compensation term $\triangle\widehat{\bm{\tilde{\Phi}}_{*}[n]}$.

\subsubsection{Gradient on the Grassmann Manifold}

Using calculus on Grassmann manifolds \cite{Absil:2009vn,Tu:2011fu},
the gradient of $\mathcal{I}(\widetilde{\bm{\Phi}})$ in (\ref{eq:gradient-mapping-F})
can be derived from $\nabla\mathcal{I}=\mathcal{P}_{\widetilde{\bm{\Phi}}}\overline{\nabla}\mathcal{I}(\widetilde{\bm{\Phi}})$,
where $\overline{\nabla}\mathcal{I}(\widetilde{\bm{\Phi}})$ is the
gradient of $\mathcal{I}(\widetilde{\bm{\Phi}})$ on the Euclidean
space and $\mathcal{P}_{\widetilde{\bm{\Phi}}}$ is to project $\mathcal{I}(\widetilde{\bm{\Phi}})$
onto the \emph{tangent space} \cite{Tu:2011fu} of $\widetilde{\bm{\Phi}}$
on the Grassmann manifold. Moreover, it is observed that, there is
no cross product of $\bm{\Phi}^{[b]}$ and $\bm{\Phi}^{[b^{'}]}$
in $\mathcal{I}(\widetilde{\bm{\Phi}})$, and hence we can compute
$F^{[b]}(\bm{\Phi}^{[b]};\mathbf{Q}^{[b]})\triangleq\nabla_{\bm{\Phi}^{[b]}}\mathcal{I}(\widetilde{\bm{\Phi}};\mathcal{Q})$
separately. As a result, the gradient $\nabla\mathcal{I}$ is given
by $F(\widetilde{\bm{\Phi}};\mathcal{Q})=\left(F^{[1]}(\bm{\Phi}^{[1]};\mathbf{Q}^{[1]}),\dots,F^{[G]}(\bm{\Phi}^{[G]};\mathbf{Q}^{[G]})\right)$,
where $F^{[b]}(\bm{\Phi}^{[b]};\mathbf{Q}^{[b]})$ is the partial
derivative w.r.t. the precoder $\bm{\Phi}^{[b]}$, $b=1,\dots,G$,
\begin{equation}
F^{[b]}(\bm{\Phi}^{[b]};\mathbf{Q}^{[b]})=\underbrace{\left[\mathbf{I}-\bm{\Phi}^{[b]}(\bm{\Phi}^{[b]\dagger}\bm{\Phi}^{[b]})^{-1}\bm{\Phi}^{[b]\dagger}\right]}_{\mbox{\scriptsize Projection }\mathcal{P}_{\bm{\Phi}^{[b]}}}\underbrace{\mathbf{Q}^{[b]}\bm{\Phi}^{[b]}}_{\mbox{\scriptsize Gradient }\nabla_{\bm{\Phi}^{[b]}}\mathcal{I}}.\label{eq:gradient}
\end{equation}

\subsubsection{Derivation of the Compensation}

\label{sub:algorithm-compensation}

Consider the parameter $\mathcal{Q}[n]$ in (\ref{prob:objective-Grassmann})
as a discrete-time sampling of the continuous-time covariance matrix
profile $\mathcal{Q}(t)$. The \emph{optimality condition} \cite{Bertsekas:1999bs,Boyd:2004kx}
of problem (\ref{prob:objective-Grassmann}) on the Grassmann manifold
$\prod_{b}\mbox{Grass}(m_{b},N_{t})$ is given by:
\begin{equation}
F(\widetilde{\bm{\Phi}}_{*};\mathcal{Q}(t))=\mathbf{0}.\label{eq:optimality-cond}
\end{equation}
Note that, since the function $\nabla\mathcal{I}(\centerdot)$ is
nonlinear, we cannot easily solve (\ref{eq:optimality-cond}) to get
$\widetilde{\bm{\Phi}}_{*}(\mathcal{Q}(t))$. However, we are only
interested in the differential $d\widetilde{\bm{\Phi}}_{*}$.

Taking the differentiation on (\ref{eq:optimality-cond}) w.r.t. $t$,
we get 
\begin{equation}
\mbox{Hess}_{\widetilde{\bm{\Phi}}_{*}}(d\widetilde{\bm{\Phi}}_{*})+F(\widetilde{\bm{\Phi}}_{*};d\mathcal{Q}(t))=\mathbf{0}\label{eq:optimality-cond-diff}
\end{equation}
where $F(\widetilde{\bm{\Phi}}_{*};d\mathcal{Q}(t))\triangleq F(\widetilde{\bm{\Phi}}_{*};\mathcal{Q}(t+dt))-F(\widetilde{\bm{\Phi}}_{*};\mathcal{Q}(t))$
is the partial differential of $F(\widetilde{\bm{\Phi}}_{*};\mathcal{Q}(t))$
on the covariance matrix profile $\mathcal{Q}(t)$, and $\mbox{Hess}_{\widetilde{\bm{\Phi}}_{*}}(d\widetilde{\bm{\Phi}}_{*})$
is the partial differential of $F(\widetilde{\bm{\Phi}}_{*};\mathcal{Q}(t))$
on $\widetilde{\bm{\Phi}}_{*}$ on the Grassmann manifold $\prod_{b}\mbox{Grass}(m_{b},N_{t})$
along the direction $d\widetilde{\bm{\Phi}}_{*}$. Note that as the
function $F(\centerdot)$ in (\ref{eq:optimality-cond}) is the gradient
of the objective function $\mathcal{I}(\widetilde{\bm{\Phi}})$ in
(\ref{prob:objective-Grassmann}), $\mbox{Hess}_{\widetilde{\bm{\Phi}}}(d\widetilde{\bm{\Phi}})$
represents the Hessian of $\mathcal{I}(\widetilde{\bm{\Phi}})$.

\mysubnote{R2-A5]}

Consider the case that the optimal solution $\widetilde{\bm{\Phi}}_{*}$
is non-degenerate, i.e., the function $F(\widetilde{\bm{\Phi}}_{*};\mathcal{Q}(t))=\mathbf{0}$
in (\ref{eq:optimality-cond}) has a unique solution over the neighborhood
of $\widetilde{\bm{\Phi}}_{*}$. By the \emph{implicit function theorem}
\cite{Tu:2011fu}, the linear equation (\ref{eq:optimality-cond-diff})
has a unique solution $d\widetilde{\bm{\Phi}}_{*}=\bm{\xi}$. Consider
that the outer precoder $\widetilde{\bm{\Phi}}$ obtained from the
previous iteration is already a good approximation of $\widetilde{\bm{\Phi}}_{*}$,
and the fact that the objective function $\mathcal{I}(\widetilde{\bm{\Phi}};\mathcal{Q})$
is decoupled on each component $\bm{\Phi}^{[b]}$, we can estimate
the differential $d\widetilde{\bm{\Phi}}_{*}$ by $\hat{\bm{\xi}}=(\bm{\hat{\xi}}^{[1]},\dots,\hat{\bm{\xi}}^{[G]})$,
where $\bm{\xi}^{[b]}$, $b=1,\dots,G$, is obtained by solving (\ref{eq:compensation-equation})
for $\bm{\xi}^{[b]}$, 
\begin{equation}
\mbox{Hess}_{\bm{\Phi}^{[b]}}(\hat{\bm{\xi}}^{[b]})+F^{[b]}(\bm{\Phi}^{[b]};d\mathbf{Q}^{[b]}(t))=\mathbf{0}.\label{eq:compensation-equation}
\end{equation}

\subsubsection{Low Complexity Calculation on Grassmann Manifold for the Compensation
Term}

Although the compensation equation (\ref{eq:compensation-equation})
is linear in the matrix variable $\hat{\bm{\xi}}$, it is a \emph{Sylvester
equation} in the general form $\mathbf{A}\hat{\bm{\xi}}+\hat{\bm{\xi}}\mathbf{B}+\mathbf{C}=\mathbf{0}$,
which is difficult to solve. However, using the property that $\hat{\bm{\xi}}$
is a point on the Grassmann manifold, we can find a low complexity
algorithm to solve the compensation equation (\ref{eq:compensation-equation}).

Consider $\bm{\Phi}^{[b]}$ are already orthonormalized. Using the
calculus on the Grassmann manifold \cite{Lundstrom:2002lh,Absil:2009vn},
the term $\mbox{Hess}_{\bm{\Phi}^{[b]}}(\hat{\bm{\xi}}^{[b]})$ can
be derived as 
\begin{equation}
\begin{array}{l}
\mbox{Hess}_{\bm{\Phi}^{[b]}}(\hat{\bm{\xi}}^{[b]})\\
=\mathcal{P}_{\bm{\Phi}^{[b]}}\left\{ \lim_{t\to0}\left[F(\bm{\Phi}_{*}^{[b]}+t\hat{\bm{\xi}}^{[b]};\mathcal{Q})-F(\bm{\Phi}_{*}^{[b]};\mathcal{Q})\right]\right\} \\
=\mathcal{P}_{\bm{\Phi}^{[b]}}\left(\mathbf{Q}^{[b]}\hat{\bm{\xi}}^{[b]}-\hat{\bm{\xi}}^{[b]}\bm{\Phi}^{[b]\dagger}\mathbf{Q}^{[b]}\bm{\Phi}^{[b]}\right)
\end{array}\label{eq:Hessian-long}
\end{equation}
Notice that $F^{[b]}(\bm{\Phi}^{[b]};d\mathbf{Q}^{[b]}(t))$ is linear
in $\bm{\Phi}^{[b]}$ (c.f. (\ref{eq:gradient})). Multiplying (\ref{eq:compensation-equation})
with a unitary matrix $\mathbf{M}$ on the right, we obtain $ $
\begin{align}
\mathcal{P}_{\bm{\Phi}^{[b]}}\bigg[\mathbf{Q}^{[b]}\big(\hat{\bm{\xi}}^{[b]}\mathbf{M}\big)-\big(\hat{\bm{\xi}}^{[b]}\mathbf{M}\big)\underbrace{\mathbf{M}^{\dagger}\bm{\Phi}^{[b]\dagger}\mathbf{Q}^{[b]}\bm{\Phi}^{[b]}\mathbf{M}}_{\bm{\Psi}}\bigg]\nonumber \\
-F^{[b]}(\bm{\Phi}^{[b]}\mathbf{M};\mathbf{Q}^{[b]})=\mathbf{0}\label{eq:compensation-equation-transformation}
\end{align}
where $\mathbf{M}$ diagonalizes $\bm{\Phi}^{[b]\dagger}\mathbf{Q}^{[b]}\bm{\Phi}^{[b]}$,
i.e., $\bm{\Phi}^{[b]\dagger}\mathbf{Q}^{[b]}\bm{\Phi}^{[b]}=\mathbf{M}\bm{\Psi}\mathbf{M}^{\dagger}$
and $\bm{\Psi}=\mbox{diag}(\beta_{1},\dots,\beta_{m_{b}})$. Let $\mathbf{Y}=\hat{\bm{\xi}}^{[b]}\mathbf{M}$.
Since ${\bf \Psi}$ is diagonal, equation (\ref{eq:compensation-equation-transformation})
can be written into $m_{b}$ parallel linear matrix equations according
to each column of $\mathbf{Y}$, 
\begin{equation}
\mathcal{P}_{\bm{\Phi}^{[b]}}\left(\mathbf{Q}^{[b]}-\beta_{i}\mathbf{I}\right)\mathbf{Y}^{i}+F^{[b]}(\bm{\Phi}^{[b]}\mathbf{M};d\mathbf{Q}^{[b]})^{i}=\mathbf{0}\label{eq:compensation-equation-decomposed}
\end{equation}
where $i=1,\dots,m_{b}$, $\mathbf{Y}^{i}$ and $F^{[b]}(\bm{\Phi}^{[b]};d\mathbf{Q}^{[b]})^{i}$
are the $i$-th ($1\leq i\leq m_{b}$) columns of $\mathbf{Y}$ and
$F^{[b]}(\bm{\Phi}^{[b]}\mathbf{M};d\mathbf{Q}^{[b]})$, respectively.
The above linear equation can be solved by the \emph{conjugate gradient}
(CG) algorithm, which only has complexity of $\mathcal{O}(N_{t}^{2})$.

\subsection{Complexity and Implementation Considerations}

\label{sub:compensation-alg-implementation}


\subsubsection{Computational Complexity}

\label{sub:complexity}

\begin{algorithm}
During each super-frame $n$, each MS $k$ in the $b$-th cell estimates
the interference covariance matrix $\mathbf{Q}_{b,k}^{[l]}[n]\triangleq\mathbb{E}\left[\mathbf{H}_{b,k}^{[l]}\mathbf{H}_{b,k}^{[l]\dagger}\right]$.
Each BS $b$ updates $\mathbf{Q}^{[b]}[n]$ according to (\ref{eq:interference-signal-covariance-BS})
and computes the new outer precoder $\bm{\Phi}^{[b]}[n+1]$ according
to the following steps:

\emph{Step 1}: Compute the compensation estimator \mysubnote{[R2-A6]}
\begin{enumerate}
\item ($2m_{b}N_{t}^{2}$ op.) Let $\triangle_{n}F^{[b]}=F^{[b]}(\bm{\Phi}_{n-1}^{[b]};\mathbf{Q}_{n}^{[b]})-F^{[b]}(\bm{\Phi}_{n-1}^{[b]};\mathbf{Q}_{n-1}^{[b]})$,
where $F^{[b]}(\centerdot)$ is given in (\ref{eq:gradient}).
\item ($2m_{b}N_{t}^{2}$ op.) Find the eigen factorization for the dimension
reduced matrix $\bm{\Phi}_{n-1}^{[b]\dagger}\mathbf{Q}^{[b]}\bm{\Phi}_{n-1}^{[b]}=\mathbf{M}\bm{\Psi}\mathbf{M}^{*}$,
where $\bm{\Psi}=\mbox{diag}(\beta_{1},\dots,\beta_{m_{b}})$.
\item ($6m_{b}N_{t}^{2}$ op.) Compute the coefficient matrix $\mathbf{A}_{i}^{[b]}=\left(\mathbf{I}-\bm{\Phi}_{n-1}^{[b]}\bm{\Phi}_{n-1}^{[b]\dagger}\right)\left(\mathbf{Q}_{n-1}^{[b]}-\beta_{i}\mathbf{I}\right)$.
\item ($4m_{b}N_{t}^{2}$ op.) Solve the equation for $\mathbf{Y}^{i}$
using CG algorithm with $N_{CG}=1$ step,
\[
\mathbf{A}_{i}^{[b]}\mathbf{Y}^{i}+(\triangle_{n}F^{[b]})^{i}=\mathbf{0},\qquad\mbox{for }1\leq i\leq m_{b}.
\]

\item Update the compensation $\bm{\Phi}_{(1)}^{[b]}\leftarrow\bm{\Phi}_{n-1}^{[b]}+\mathbf{Y}\mathbf{M}^{*}$.
\end{enumerate}
\emph{Step 2}: ($2m_{b}N_{t}^{2}$ op.) Compute the search direction
$\bm{\eta}^{[b]}=\mathbf{Q}_{n}^{[b]}\bm{\Phi}_{(1)}^{[b]}$, for
$b=1,\dots,G$.

\emph{Step 3}: Update $\bm{\Phi}_{n+1}^{[b]}\leftarrow\mbox{qr}\left(\bm{\Phi}_{(1)}^{[b]}-\gamma\bm{\eta}^{[b]}\right)$,
where $\gamma>0$ is the step size and $\mbox{qr}(\mathbf{A})$ denotes
the Gram-Schmidt procedure for the orthogonalization of $\mathbf{A}$.

\caption{\label{alg:compensation} Compensation algorithm for the outer precoding}
\end{algorithm}

The computational complexity of the proposed compensation algorithm
(\ref{eq:compensation-algorithm}) is mainly contributed by the gradient
term $F(\widetilde{\bm{\Phi}}[n];\mathcal{Q}[n])$ and the compensation
term $\triangle\widehat{\bm{\tilde{\Phi}}_{*}[n]}$ in (\ref{eq:compensation-algorithm}).
The gradient term requires $2m_{b}N_{t}^{2}$ (omitting the small
order terms) arithmetic operations (addition, multiplication, etc.).
The compensation term requires solving the linear equations in (\ref{eq:compensation-equation-decomposed})
with the CG algorithm. Note that, as the CG algorithm has a fast convergence
rate and the norm of $F(\bm{\widetilde{\Phi}};d\mathcal{Q})$ is usually
small (since $d\mathcal{Q}$ is small due to the slow time-varying
property of the covariance matrix $\mathbf{Q}^{[b]}(t)$), computing
only $N_{CG}=1$ step (requires $4m_{b}N_{t}^{2}$ operations) to
obtain $\hat{\bm{\xi}}^{[b],i}$ is sufficient to yield a good compensation
$\triangle\widehat{\bm{\tilde{\Phi}}_{*}(n)}$. Therefore, the proposed
compensation algorithm (summarized in Algorithm 1) has a total computational
complexity of around $16m_{b}N_{t}^{2}$ operations. As we discussed
in Section \ref{sub:motivation-two-tier}, we usually have $m_{b}\ll N_{t}$
for massive MIMO channels, and therefore the complexity of the proposed
algorithm is substantially lower than $\mathcal{O}(N_{t}^{3})$ of
the brute force computing of Theorem \ref{thm:solution-outer-precoder}
using SVD \cite{Golub:2012kx}.

\subsubsection{Implementation Considerations}

\label{sub:implementation-issues}

\begin{figure}
\begin{centering}
\includegraphics[width=0.7\columnwidth]{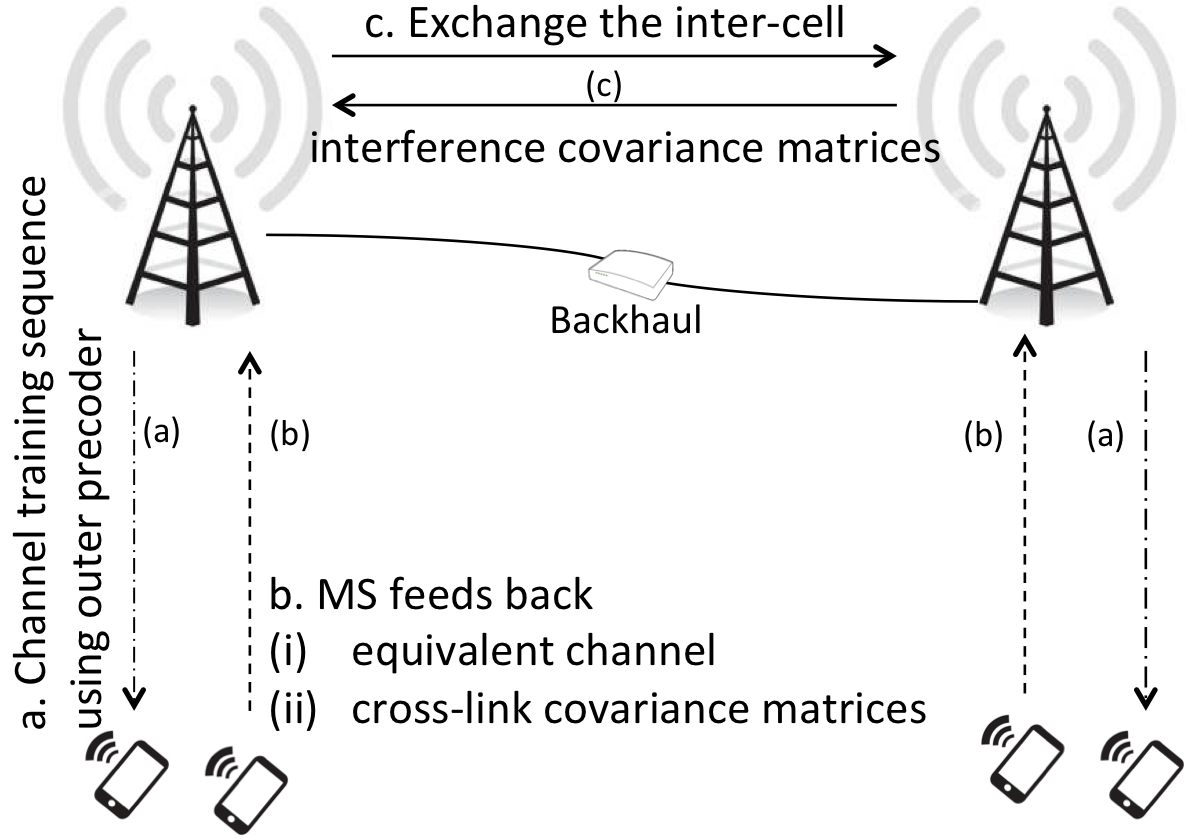}
\par\end{centering}

\caption{\label{fig:signaling} Diagram of the signaling in the multi-cell
massive MIMO system.}
\end{figure}

Fig. \ref{fig:signaling} gives a diagram of the associated signaling
for the two-tier precoding. In stage (a), each BS $b$ broadcasts
channel training sequences using the outer precoder $\bm{\Phi}^{[b]}$
at each subframe. In stage (b), each MS feeds back the low dimensional
equivalent channel at each subframe and full dimension ($N_{t}\times N_{t})$
interference covariance matrices $\{\mathbf{Q}_{b,k}^{[l]}\}$ only
at the end of each super-frame. In stage (c), BSs exchange the covariance
matrix profile $\{\mathbf{Q}_{b,1}^{[l]}\}$ for each MS cluster through
the backhaul only at the end of each super-frame. As a result, the
pilot symbols for channel estimation, the CSI feedback overhead and
the signaling over the backhaul have been greatly reduced in the massive
MIMO system.

\section{Convergence Analysis of the Outer Precoding Algorithm }

\label{sec:convergence-analysis}


In this section, we analyze the tracking performance of the proposed
iterative outer precoder tracking algorithms under time-varying channels.
We are interested in whether algorithm (\ref{eq:compensation-algorithm})
will converge to the global optimal solution of the problem in (\ref{eq:algorithm-outer-precoder}).
However, since the problem in (\ref{eq:algorithm-outer-precoder})
is non-convex and there are multiple\emph{ }stationary points for
the algorithm, existing techniques \cite{Chen2012:Saddle,Chen13-two}
for the convergence analysis under time-varying channels cannot be
applied. In general, we shall address the following challenge:\\
\framebox{\begin{minipage}[t]{1\columnwidth}%
\textbf{Challenge 2}: Analyze the convergence behavior for the outer
precoder tracking algorithm under the time-varying massive MIMO channel,
despite the optimization problem being non-convex.%
\end{minipage}}\medskip{}

Towards this end, we extend the analysis framework in \cite{Chen2012:Saddle,Chen13-two}
and obtain the results of  the algorithm tracking performance by analyzing
an equivalent continuous-time virtual dynamic system (VDS), which
models the behavior of the algorithm iteration. Please refer to Appendix
\ref{app:VDS} for details.

\subsection{Convergence under Static Channel Covariance}

\label{sub:convergence-static}

When the channel covariance matrices $\mathbf{Q}^{[b]}$ are static,
the compensation term in iteration (\ref{eq:compensation-algorithm})
is always zero. Hence, the compensation algorithm (\ref{eq:compensation-algorithm})
degenerates to a pure gradient descent algorithm (\ref{eq:gradient-algorithm}).
To establish the convergence results, we first derive the following
uniqueness property for the algorithm. 
\begin{lyxLemQED}
[Uniqueness of Global Optimal Point]\label{lem:unique-stable-equilibrium}
Suppose under a given $\mathcal{Q}=(\mathbf{Q}^{[1]},\dots,\mathbf{Q}^{[G]})$,
the covariance matrix $\mathbf{Q}^{[b]}$ has distinct $m_{b}$-th
and $(m_{b}+1)$-th smallest eigenvalues $\lambda_{m_{b}}^{[b]}\neq\lambda_{m_{b}+1}^{[b]}$
for each $b$. Then there is only one global optimal stationary point
for the iteration (\ref{eq:compensation-algorithm}). 
\end{lyxLemQED}
\begin{proof}
Please refer to Appendix \ref{app:VDS} for the proof.
\end{proof}

Based on Lemma \ref{lem:unique-stable-equilibrium}, we shall establish
the global convergence result below. 
\begin{lyxThmQED}
\label{thm:global-convergence} \emph{(Global Convergence under Static
Channel Covariance)} There exists $\gamma_{0}>0$, such that under
the distinct eigenvalue condition in Lemma \ref{lem:unique-stable-equilibrium}
and choosing step size $0<\gamma<\gamma_{0}$, the proposed algorithm
converges to the global optimal solution $\widetilde{\bm{\Phi}}_{*}(\mathcal{Q})$.
\end{lyxThmQED}
\begin{proof}
Please refer to Appendix \ref{app:VDS} for the proof.
\end{proof}

The above theorem concludes that, although the original outer precoding
problem (\ref{eq:algorithm-outer-precoder}) is non-convex, the proposed
algorithm is guaranteed to converge to the global optimal solution
under static channel covariance.

\begin{remrk}
[Global Convergence of Non-Convex Problem] As we pointed out in
Section \ref{sub:formulation-grassmann-manifold}, problem (\ref{eq:algorithm-outer-precoder})
is non-convex. Yet, after the problem reformation, the new problem
in (\ref{prob:objective-Grassmann}) on the manifold has the following
structure: there is only one maximum point (attractive) among all
the other KKT points (repulsive) as illustrated in Fig. \ref{fig:saddle-points}.
As a result, the iterative algorithm trajectory will converge to the
maximum point almost surely. 

\begin{figure}
\begin{centering}
\includegraphics[width=0.75\columnwidth]{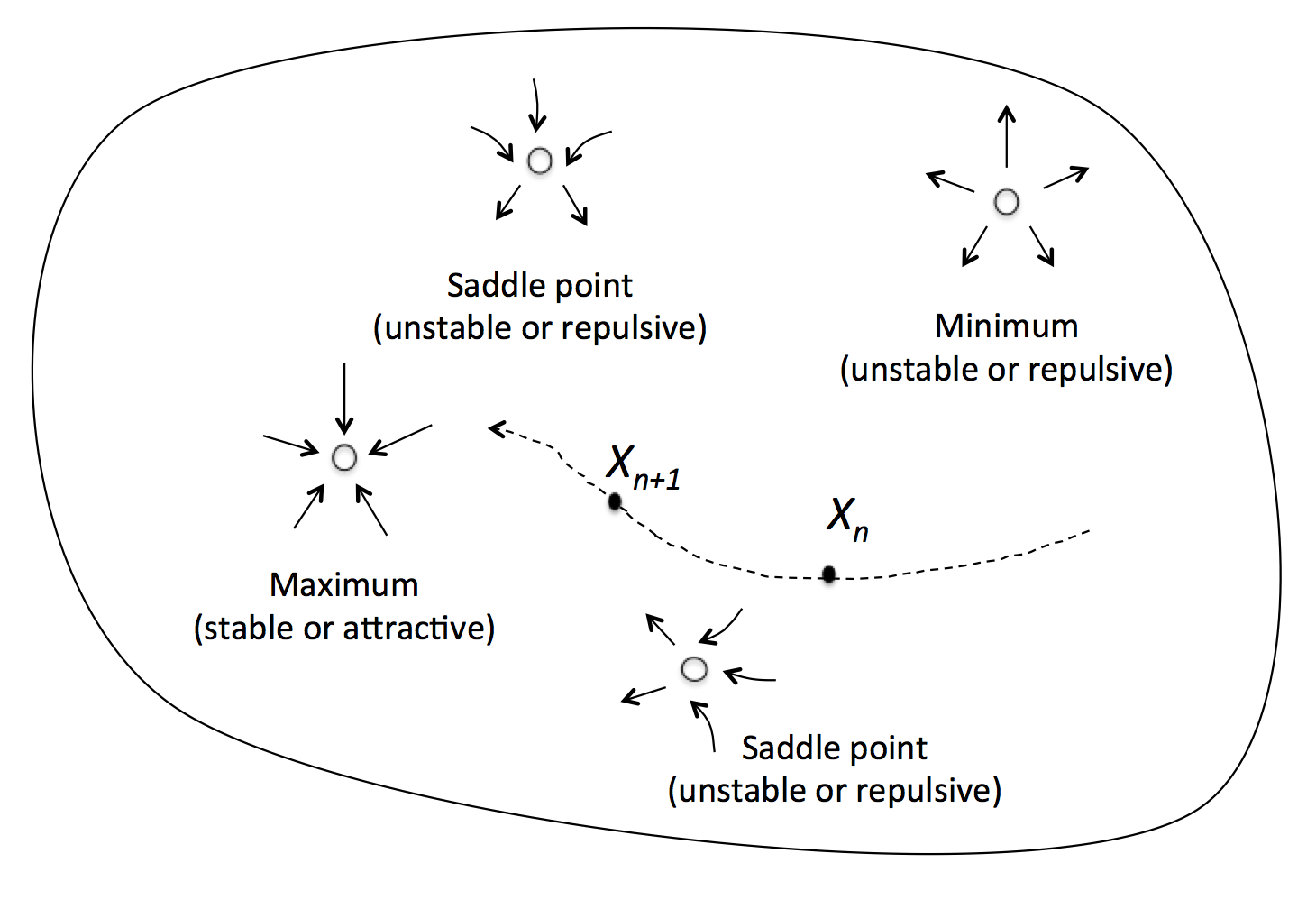}
\par\end{centering}

\caption{\label{fig:saddle-points} An illustration of the KKT points of problem
(\ref{prob:objective-Grassmann}) on the Grassmann manifold. Except
for the global maximum point, other KKT points (stationary points
or minimum point) are unstable (repulsive).}
\end{figure}

\end{remrk}

\subsection{Convergence under Time-Varying Channel Covariance}

\label{sub:convergence-time-varying}

We now study the case under time-varying channel covariance. Under
time-varying channels, the global optimal solution $\widetilde{\bm{\Phi}}^{*}(t)$
is also time varying in similar timescale as the algorithm iteration
(\ref{eq:compensation-algorithm}) and hence, it is not clear if the
iterate $\widetilde{\bm{\Phi}}[n]$ can converge to $\widetilde{\bm{\Phi}}^{*}(t_{n})$. 

To analyze the tracking performance of the outer precoder iterations
in (\ref{eq:compensation-algorithm}), we approximate the discrete-time
iterations $\widetilde{\bm{\Phi}}[n]$ with the following continuous
time iterations%
\footnote{The iteration $\widetilde{\bm{\Phi}}[n]$ of (\ref{eq:compensation-algorithm})
is a discretization of the compensated virtual dynamic system $\widetilde{\bm{\Phi}}^{c}(t)$
at $t=n\tau T_{s}$ (for example, by replacing $dt$ in (\ref{eq:compensation-alg-flow})
with $\gamma_{n}\approx\triangle t_{n}$ in (\ref{eq:compensation-algorithm})). %
} $\widetilde{\bm{\Phi}}^{c}(t)$, which is defined as the solution
of the following differential equations:

\begin{eqnarray}
d\widetilde{\bm{\Phi}}^{c} & = & F(\widetilde{\bm{\Phi}}^{c};\mathcal{Q}(t)dt+\widehat{d\widetilde{\bm{\Phi}}_{*}},\qquad\widetilde{\bm{\Phi}}^{c}(0)=\widetilde{\bm{\Phi}}_{0}\label{eq:compensation-alg-flow}\\
\mathbf{0} & = & \mbox{Hess}_{\widetilde{\bm{\Phi}}^{c}}(d\widetilde{\bm{\Phi}}^{c})+F(\widetilde{\bm{\Phi}}^{c};d\mathcal{Q}(t)).\label{eq:continuous-time-compensation-flow}
\end{eqnarray}

We evaluate the tracking behavior of the outer precoder iteration
$\widetilde{\bm{\Phi}}^{c}(t)$ in the following.
\begin{lyxThmQED}
\label{thm:convergence-compensation} \emph{(Convergence of the Outer
Precoder Iteration in Time-varying Channels)} Assume the distinct
eigenvalue condition in Lemma \ref{lem:unique-stable-equilibrium}.
In addition, suppose the largest eigenvalue $\lambda_{\max}$ of $\mathbf{Q}^{[b]}$
is bounded w.p.1 for all $b$. Then there exists $\delta>0$, such
that for $\|\widetilde{\bm{\Phi}}^{c}(0)-\widetilde{\bm{\Phi}}_{*}(0)\|_{F}<\delta$,
we have $\|\widetilde{\bm{\Phi}}^{c}(t)-\widetilde{\bm{\Phi}}_{*}(t)\|_{F}\to0$,
w.p.1, as $t\to\infty$. 
\end{lyxThmQED}
\begin{proof}
Please refer to Appendix \ref{app:pf-thm-convergence-compensation}
for the proof.
\end{proof}

The result in Theorem \ref{thm:convergence-compensation} implies
that perfect tracking of the outer precoder in time-varying channels
is possible if the initial iterate $\widetilde{\bm{\Phi}}[0]$ is
sufficiently close to the global optimal point $\widetilde{\bm{\Phi}}_{*}(0)$.

\section{Numerical Results}

\label{sec:numerical}

\begin{figure}
\begin{centering}
\includegraphics[width=0.9\columnwidth]{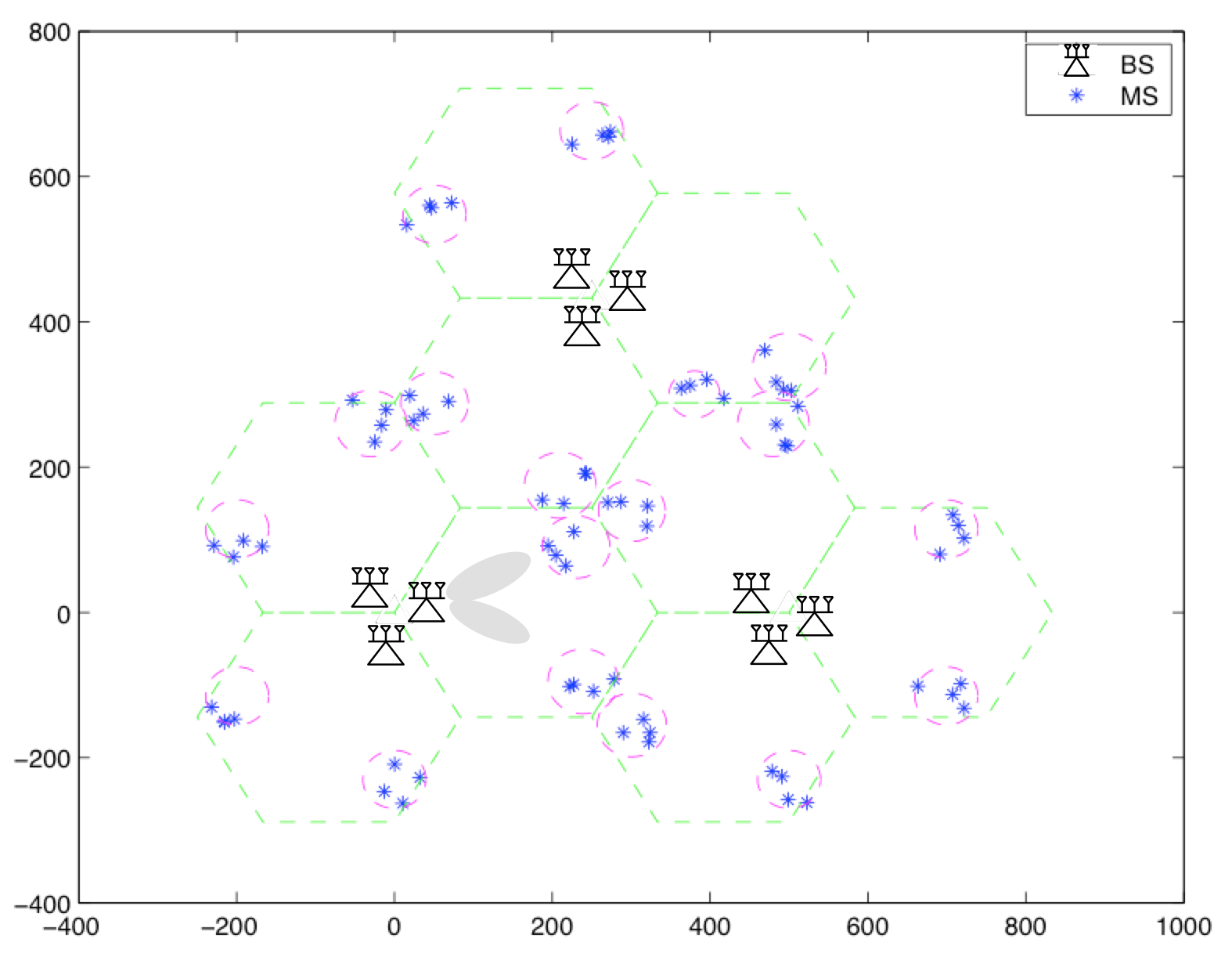}
\par\end{centering}

\caption{\label{fig:topology} Topology of a cellular network with clusters
of users.}
\end{figure}

We consider a cellular network with $G=9$ cells, where each cell
has 2 clusters and each cluster has $K=4$ users. Fig. \ref{fig:topology}
illustrates a realization of the network topology. The inter-site
distance is $500$ m. The large-scale propagation follows the outdoor
evaluation methodology in LTE standard \cite{TR36814} with pathloss
exponent 2.6. Each BS is equipped with $N_{t}=48$ antennas and each
user is equipped with $N_{r}=1$ antenna. We generate the massive
MIMO channels according to (\ref{eq:two-timescale-channel-model}),
where the transmit correlation matrices are specified by (\ref{eq:chann-spatial-correlation})
and the AS parameter is modeled as $\triangle_{b,k}^{[l]}=20$ deg.
Moreover, the small timescale channel variation $\mathbf{H}_{k}^{\omega}(j)$
in (\ref{eq:two-timescale-channel-model}) is modeled by the widely
used autoregressive (AR) model \cite{Baddour01} given by $\mathbf{H}_{k}^{\omega}(j)=\theta\mathbf{H}_{k}^{\omega}(j-1)+\sqrt{1-\theta^{2}}\mathbf{W}$,
where $\mathbf{W}$ is a standard complex Gaussian matrix, $\theta=J_{0}(2\pi f_{d}\tau)$
is the temporal correlation coefficient, $J_{0}(\centerdot)$ is the
zero-th order Bessel function, $f_{d}$ is the maximum Doppler frequency,
and $\tau=1$ is the subframe duration. The length of the super-frame
is $T_{s}=100$. The noise is normalized as equal to the smallest
direct link power gain. 

We consider the following baselines: \textbf{Baseline 1 (One-tier
coordinated MIMO using ZF} \textbf{\cite{Yoo:2006fk})}: In each subframe,
full CSI is used to compute the precoder, which zero-forces both the
inter-cell and intra-cell interference.\textbf{ Baseline 2 (Two-tier
precoding using the BD algorithm in \cite{Adhikary:2012vn})}: Two-tier
precoding strategy in \cite{Adhikary:2012vn} is applied, where the
outer precoder is computed by the BD algorithm in \cite{Adhikary:2012vn}.\textbf{
Baseline 3 (Two-tier precoding with conventional gradient algorithm
for the outer precoder \cite{Poon:2003zr})}: The two-tier precoding
strategy (equations (\ref{eq:algorithm-outer-precoder}), (\ref{eq:algorithm-decoder})
and (\ref{eq:algorithm-inner-precoder})) is applied, where the solution
of the outer precoders given in Theorem \ref{thm:solution-outer-precoder}\textbf{
}are computed iteratively using the gradient algorithm in \cite{Poon:2003zr}.

Note that Baseline 1 suffers from implementation challenges in massive
MIMO systems, such as huge pilot symbols and feedback overhead, and
real-time global CSI sharing as discussed in Section \ref{sec:intro}.
Hence it serves as performance benchmark only.

\subsection{Throughput Performance}

\begin{figure}
\begin{centering}
\includegraphics[width=1\columnwidth]{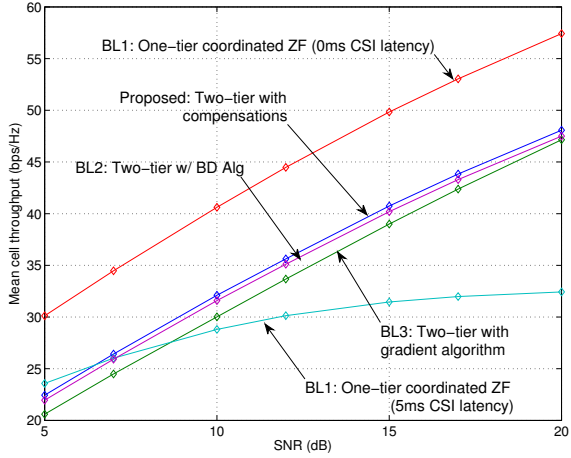}
\par\end{centering}

\caption{\label{fig:throghput-SNR} The per cell throughput versus the per
BS transmit power under MS speed 10 km/h. }
\end{figure}

Fig. \ref{fig:throghput-SNR} shows the per cell throughput versus
the per BS transmit power under MS speed 10 km/h. The one-tier cooperative
ZF scheme (Baseline 1) achieves the highest data rate when there is
no signaling latency for the BSs to exchange global CSI over the backhaul.
However, the performance of Baseline 1 is very sensitive to the signaling
latency and its performance degrades significantly when 5 ms backhaul
latency is considered%
\footnote{As a benchmark, the X2 interface in e-Node B of LTE systems usually
induce 10-20 ms latency \cite{TR36814}.%
}. On the other hand, the performance of the two tier precoding schemes
(Baseline 2, Baseline 3 and proposed scheme) are robust to signaling
latency, as they do not require instantaneous global CSI. The proposed
scheme achieves slightly better performance compared with Baseline
2 but with substantially lower complexity (Table \ref{tab:complexity}). 

\begin{figure}
\begin{centering}
\includegraphics[width=1\columnwidth]{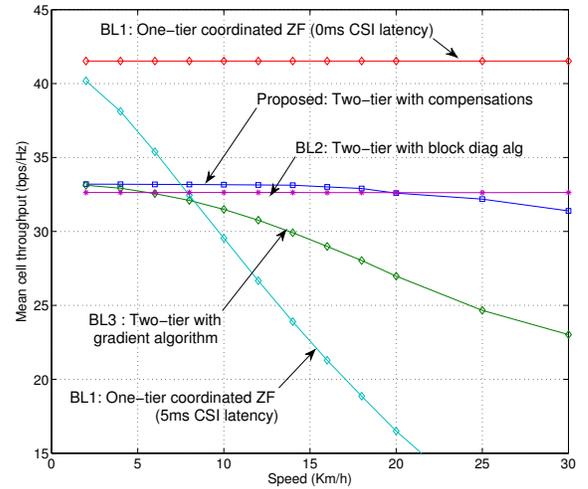}
\par\end{centering}

\caption{\label{fig:throghput-velocity} The per cell throughput versus the
MS mobility under per BS transmit power $P=10$ dB. }
\end{figure}

Fig. \ref{fig:throghput-velocity} shows the per cell throughput versus
the MS speed under per BS transmit power budget $P=10$ dB. Similarly,
the proposed scheme with compensation achieves good performance but
with substantially lower complexity. In addition, it significantly
outperforms Baseline 3 at high MS speed. This confirms the superior
tracking capability of the proposed compensation algorithm under time-varying
channels. As a comparison, the throughput performance of Baseline
1 drops quickly when increasing the MS speed under $5$ ms backhaul
latency.

\subsection{Feedback Loading and Complexity}

Table \ref{tab:CSI-feedback-amount} shows numerical examples of CSI
feedback amount and signaling loading in terms of number of complex
numbers per cell per subframe following the discussion in Section
\ref{sub:implementation-issues}. We assume each cell only needs to
exchange CSI to $3$ neighboring cells. Baseline 1 requires a high
feedback cost and signaling loading. Whereas, the proposed scheme
has significantly lowered the CSI feedback overhead among BSs in massive
MIMO systems.

\begin{table}
\begin{centering}
\begin{tabular}{|c|c|c|c|c|}
\hline 
\multirow{2}{*}{$(N_{t,}K)$} & \multicolumn{2}{c|}{Feedback amount } & \multicolumn{2}{c|}{Signaling loading}\tabularnewline
\cline{2-5} 
 & BL 1  & BL 2, 3 \& Prop & BL 1 & BL 2, 3 \& Prop\tabularnewline
\hline 
\hline 
 & $N_{t}N_{r}K$ & $N_{t}^{2}/T_{s}+N_{r}K^{2}$ & $3N_{t}N_{r}K$ & $3N_{t}^{2}/T_{s}$\tabularnewline
\hline 
$(24,8)$ & 384 & 134 & 1,152 & 17\tabularnewline
\hline 
$(48,8)$ & 768 & 151 & 2,304 & 69\tabularnewline
\hline 
$(100,30)$ & 6,000 & 1,900 & 18,000 & 300\tabularnewline
\hline 
\end{tabular}
\par\end{centering}

\caption{\label{tab:CSI-feedback-amount} Average CSI feedback amount and signaling
loading in terms of number of complex numbers per cell per subframe,
where $N_{r}=2$, the dimension of the outer precoder is $m=K$, $T_{s}=100$,
and one cluster per cell.}
\end{table}

Table \ref{tab:complexity} summarizes the computational complexity%
\footnote{The major complexity of these algorithms consists for the complexity
of pseudo inverse (from ZF solution) \cite{Moller98} and the complexity
of eigen analysis.%
} in terms of the millions of complex multiply-accumulate (MCMA) operations
per super-frame and the corresponding computational time%
\footnote{The computational time is estimated by using a TI-TMS320DM642 DSP,
which can execute up to 4800 million instructions per second (MIPS).
Operation overheads such as memory loading are not counted.%
} for computing the outer precoder for one cluster under a two-tier
precoding strategy. The proposed algorithm has a much lower complexity
than the BD algorithm (Baseline 2) and the scheme using the brute
force computation of Theorem \ref{thm:solution-outer-precoder}.

\begin{table}
\begin{centering}
\begin{tabular}{|c|c|c|c|}
\hline 
\multirow{3}{*}{$(N_{t,}K)$} & \multicolumn{3}{c|}{MCMA \& time (ms)}\tabularnewline
\cline{2-4} 
 & BD (BL 2)  & SVD & Proposed\tabularnewline
\cline{2-4} 
 & $21N_{t}^{3}+21(N_{t}-r^{*})^{3}$ & $21N_{t}^{3}$ \cite{Golub:2012kx} & $16mN_{t}^{2}$\tabularnewline
\hline 
$(24,8)$ & 0.37 (0.078) & 0.29 (0.06) & 0.07 (0.02)\tabularnewline
\hline 
$(48,8)$ & 3.67 (0.76) & 2.3 (0.48) & 0.29 (0.06)\tabularnewline
\hline 
$(100,8)$ & 37 (7.8) & 21 (4.4) & 1.3 (0.27)\tabularnewline
\hline 
\end{tabular}
\par\end{centering}

\caption{\label{tab:complexity} Rough calculation of the computational complexity
in terms of millions of complex multiply-accumulate (MCMA) operations
per super-frame and the corresponding computational time in milliseconds
using TI-TMS320DM642 DSP for computing the outer precoder for one
cluster. The dimension of the outer precoder for the proposed scheme
is chosen as $m=r^{*}=K$.}
\end{table}

\section{Conclusions}

\label{sec:conclustion}

In this paper, we propose a low complexity compensation algorithm
for tracking the outer precoder under the two-tier precoding in massive
MIMO systems and time-varying channels. The two-tier precoding scheme
tries to combat various implementation challenges raised in massive
MIMO systems, namely, the huge pilot symbols and feedback overhead,
real-time global CSI requirement, large number of RF chains and high
computational complexity. In particular, to reduce the computational
complexity for the outer precoder, we propose an iterative algorithm
which is derived by solving an optimization problem formulated on
the Grassmann manifold, and its tracking performance is enhanced by
leveraging a compensation technique to offset the time variation of
the optimal solution. We show with analytical results that, under
some mild conditions, perfect tracking of the outer precoder is possible.
The numerical results also confirm the superior performance advantage
of the two-tier precoding with the proposed compensation algorithm.

\appendices

\section{Proof of Theorem \ref{thm:solution-outer-precoder}}

\label{app:pf-thm-the-outer-precoder-solution}

Denote $\widetilde{\bm{\Phi}}=(\bm{\Phi}^{[1]},\dots,\bm{\Phi}^{[G]})$
as the subspace precoder profile for all the BSs. The objective function
in the outer precoding problem in (\ref{eq:algorithm-outer-precoder})
can be written as 
\begin{align*}
\mathcal{I}(\widetilde{\bm{\Phi}}) & \triangleq\sum_{b=1}^{G}\sum_{l\neq b}\sum_{k=1}^{K_{l}}\mbox{tr}\left[\mathbb{E}\left(\mathbf{H}_{b,k}^{[l]}\bm{\Phi}^{[l]}\bm{\Phi}^{[l]\dagger}\mathbf{H}_{b,k}^{[l]\dagger}\right)\right]\\
 & \qquad-w\sum_{b=1}^{G}\sum_{k=1}^{K_{l}}\mbox{tr}\left[\mathbb{E}\left(\mathbf{H}_{b,k}^{[b]}\bm{\Phi}^{[b]}\bm{\Phi}^{[b]\dagger}\mathbf{H}_{b,k}^{[b]\dagger}\right)\right]\\
 & =\sum_{b=1}^{G}\mbox{tr}\bigg\{\bm{\Phi}^{[b]\dagger}\bigg[\sum_{l\neq b}\sum_{k=1}^{K_{l}}\mathbb{E}\left(\mathbf{H}_{l,k}^{[b]\dagger}\mathbf{H}_{l,k}^{[b]}\right)\\
 & \qquad\qquad\qquad-w\sum_{k=1}^{K_{b}}\mathbb{E}\left(\mathbf{H}_{b,k}^{[b]\dagger}\mathbf{H}_{b,k}^{[b]}\right)\bigg]\bm{\Phi}^{[b]}\bigg\}\\
 & =\sum_{b=1}^{G}\mbox{tr}\left\{ \bm{\Phi}^{[b]\dagger}\mathbf{Q}^{[b]}\bm{\Phi}^{[b]}\right\} 
\end{align*}
which is equivalent to solving the following minimization problem

\begin{eqnarray}
\min_{\{\bm{\Phi}^{[b]\dagger}\bm{\Phi}^{[b]}=\mathbf{I}_{m_{b}}\}} & \mathcal{I}(\widetilde{\bm{\Phi}})=\sum_{b=1}^{G}\mbox{tr}\left\{ \bm{\Phi}^{[b]\dagger}\mathbf{Q}^{[b]}\bm{\Phi}^{[b]}\right\} .\label{eq:algorithm-outer-precoder-reformulation}
\end{eqnarray}

Applying eigenvalue decomposition (EVD) to $\mathbf{Q}^{[b]}$, we
get $\mathbf{Q}^{[b]}=\mathbf{W}^{[b]}\bm{\Lambda}^{[b]}\mathbf{W}^{[b]\dagger}$.
Thus due to the unitary constraint $\bm{\Phi}^{[b]\dagger}\bm{\Phi}^{[b]}=\mathbf{I}_{m_{b}}$,
the optimal solution is given by the $m_{b}$ columns of $\mathbf{W}^{[b]}$
corresponding to the $m_{b}$ smallest diagonal elements of $\bm{\Lambda}^{[b]}$.

\section{Proof of Theorem \ref{thm:Achievable-DoF}}

\label{app:pf-lem-achievable-DoF}

We first consider the outer bound of the DoF $\Gamma_{\mbox{\scriptsize one}}$
under the one-tier IA. Suppose we allow receiver cooperation in each
cluster. As a result, the MIMO interference broadcast channel becomes
a $G$-pair $KN_{r}\times N_{t}$ interference channel, and we denote
the $KN_{r}\times N_{t}$ concatenated channel as $\widetilde{\mathbf{H}}_{b}^{[l]}=[\mathbf{H}_{b,1}^{[l]\dagger},\dots,\mathbf{H}_{b,K}^{[l]\dagger}]^{\dagger}.$

\mysubnote{R2-A2}

Applying EVD to the transmit covariance matrix, we get $\mathbf{T}_{b}^{[l]}=\mathbf{U}_{T,b}^{[l]}\bm{\Lambda}_{T,b}^{[l]}\mathbf{U}_{T,b}^{[l]\dagger}=\sum_{j=1}^{\Upsilon}\lambda_{b,j}^{[l]}\mathbf{u}_{T,b,j}^{[l]}\mathbf{u}_{T,b,j}^{[l]\dagger},$
where $\bm{\Lambda}_{T,b}^{[l]}$ is a diagonal matrix with diagonal
elements $\{\lambda_{b,j}^{[l]}\}_{j=1}^{\Upsilon}$ sorted in a descent
order and $\mathbf{u}_{T,b}^{[l],j}$ is the $j$-th column of $\mathbf{U}_{T,b}^{[l]}$.
From the channel model in (\ref{eq:two-timescale-channel-model}),
we have $\mathbf{H}_{b,k}^{[l]}=\mathbf{H}^{w}\sum_{j=1}^{\Upsilon}\sqrt{\lambda_{b,j}^{[l]}}\mathbf{u}_{T,b,j}^{[l]}\mathbf{u}_{T,b,j}^{[l]\dagger}=\sum_{j=1}^{\Upsilon}\hat{\mathbf{H}}_{j}^{w}\mathbf{u}_{T,b,j}^{[l]\dagger}$,
where $\hat{\mathbf{H}}_{j}^{w}=\sqrt{\lambda_{b,j}^{[l]}}\mathbf{H}^{w}\mathbf{u}_{T,b,j}^{[l]}$
and each row of $\mathbf{H}_{b,k}^{[l]}$ is a linear combination
of the $\Upsilon$ vectors $\{\mathbf{u}_{T,b,j}^{[l]}\}_{j=1}^{\Upsilon}$.
As a result, the concatenated matrix $\widetilde{\mathbf{H}}_{b}^{[l]}$
has rank at most $\min\{\Upsilon,KN_{r}\}$. Therefore, using the
result in \cite[Theorem 1]{Chae:2011vn}, the $G$-pair $KN_{r}\times N_{t}$
rank deficient interference channel has a per-cell DoF $\min\{\Upsilon,KN_{r}\}$,
and thus, we obtain an outer bound $\Gamma_{\mbox{\scriptsize one}}\leq\min\{\Upsilon,KN_{r}\}$. 

\mysubnote{R1-A3.e}

We now derive the inner bound (achievability) of the DoF $\Gamma_{\mbox{\scriptsize two}}$
under the two-tier precoding. Construct a principal matrix $\mathbf{\hat{U}}_{T,b}^{[l]}$
for the covariance matrix $\mathbf{T}_{b}^{[l]}$ of each link by
extracting the $\Upsilon$ major eigen components, i.e., $\mathbf{\hat{U}}_{b}^{[l]}=[\mathbf{u}_{T,b,1}^{[l]}\,\mathbf{u}_{T,b,2}^{[l]}\,\dots\,\mathbf{u}_{T,b,\Upsilon}^{[l]}]$,
where $\{\mathbf{u}_{T,b,j}^{[l]}\}$ are the major eigen modes of
the transmit covariance matrix from BS $l$ to the clustered users
in cell $b$. Construct the principal signal matrix for each BS $b$
as $\hat{\mathbb{U}}_{b}=[\hat{\mathbf{U}}_{b}^{[1]}\,\dots\,\hat{\mathbf{U}}_{b}^{[G]}]^{\dagger}$
with dimension $(G\Upsilon)\times N_{t}$. Since all the elements
of $\hat{\mathbb{U}}_{b}$ are independent and continuous distributed,
$\hat{\mathbb{U}}_{b}$ has full row rank $G\Upsilon$ under the condition
that $N_{t}\geq G\Upsilon$. As a result, we can find a pseudo-inverse
$\hat{\mathbb{U}}_{b}^{\ddagger}$ of $\hat{\mathbb{U}}_{b}$, such
that $\hat{\mathbb{U}}_{b}\hat{\mathbb{U}}_{b}^{\ddagger}=\mathbf{I}_{G\Upsilon}$.
This implies that, there exists a $N_{t}\times\Upsilon$ matrix $\bm{\mathbf{\Phi}}_{\Upsilon}^{[b]}$
that $\mathbf{\hat{U}}_{b}^{[b]}\bm{\mathbf{\Phi}}_{\Upsilon}^{[b]}=\mathbf{I}_{\Upsilon}$,
$\forall b$, and $\mathbf{\hat{U}}_{b}^{[l]}\bm{\mathbf{\Phi}}_{\Upsilon}^{[b]}=\mathbf{0}$,
$\forall l\neq b$. As a result, we have 
\begin{equation}
\mbox{rank}(\mathbf{T}_{b}^{[b]}\bm{\mathbf{\Phi}}_{\Upsilon}^{[b]})=\Upsilon,\qquad\mbox{and }\mathbf{T}_{b}^{[l]}\bm{\mathbf{\Phi}}_{\Upsilon}^{[b]}=\mathbf{0},\forall l\neq b.\label{eq:app-IA-condition}
\end{equation}
Therefore, with probability 1, we have $\mbox{rank}(\mathbf{H}_{b,k}^{[b]}\bm{\mathbf{\Phi}}_{\Upsilon}^{[b]})=\min\{N_{r},\Upsilon\}$,
$\forall b$, and $\mathbf{H}_{b,k}^{[l]}\bm{\mathbf{\Phi}}_{\Upsilon}^{[l]}$,
$\forall l\neq b$. Hence, the concatenated matrix $\widetilde{\mathbf{H}}_{b}^{[b]}\bm{\Phi}_{\Upsilon}^{[b]}$
has row rank $\min\{\Upsilon,KN_{r}\}$, and by using transmit ZF,
$\min\{\Upsilon,KN_{r}\}$ DoF can be achieved per cell.

Choose the dimension of the outer precoders $\{\bm{\Phi}^{[b]}\}_{b=1}^{G}$
as $N_{t}\times m$, where $m=\min\{\Upsilon,KN_{r}\}$. Denote $\widetilde{\bm{\Phi}}=(\bm{\Phi}^{[1]},\dots,\bm{\Phi}^{[G]})$.
 Consider $w=0$; then the set of optimal solutions $\widetilde{\bm{\Phi}}_{*}\big|_{w=0}$
must satisfy the nulling condition in (\ref{eq:app-IA-condition})
and the optimal value of (\ref{eq:algorithm-outer-precoder}) is $0$.
Due to the continuity of the objective function (quadratic forms),
there exists a $\epsilon$-neighborhood $\mathcal{N}_{0}(\widetilde{\bm{\Phi}}_{*})$
of some $\widetilde{\bm{\Phi}}_{*}\big|_{w=0}$, such that we can
find a $\widetilde{\bm{\Phi}}\in\mathcal{N}_{0}(\widetilde{\bm{\Phi}}_{*})$
to satisfy all the subspace alignment conditions in (\ref{eq:app-IA-condition}).
Hence, choosing $w$ sufficiently small, e.g., $w=\mathcal{O}(\frac{1}{\rho})$
under high SNR $\rho$, the optimal solution $\widetilde{\bm{\Phi}}_{*}$
to (\ref{eq:algorithm-outer-precoder}) satisfies the condition (\ref{eq:app-IA-condition}).
As a result, it guarantees that $\Gamma_{\mbox{\scriptsize two}}\geq\min\{\Upsilon,KN_{r}\}$.
As $\Gamma_{\mbox{\scriptsize two}}\leq\Gamma_{\mbox{\scriptsize one}}$,
we must have $\Gamma_{\mbox{\scriptsize two}}=\Gamma_{\mbox{\scriptsize one}}=\min\{\Upsilon,KN_{r}\}$.

\section{Convergence via Virtual Dynamic System Modeling}

\label{app:VDS}

\subsection{The Virtual Dynamic System (VDS) Modeling and Proof of Lemma \ref{lem:unique-stable-equilibrium}}

We first model the algorithm iteration into a continuous-time virtual
dynamic system defined as follows. 
\begin{lyxDefQED}
[Continuous-time Virtual Dynamic System] The continuous-time virtual
dynamic system (VDS) is defined by the state trajectory $\widetilde{\bm{\Phi}}^{c}(t)$,
which is the solution to the following differential equation 
\begin{equation}
d\widetilde{\bm{\Phi}}^{c}=-F(\widetilde{\bm{\Phi}}^{c};\mathcal{Q}(t))dt,\qquad\widetilde{\bm{\Phi}}^{c}(0)=\widetilde{\bm{\Phi}}[0]\label{eq:continuous-time-dynamic-system}
\end{equation}
where $\widetilde{\bm{\Phi}}[0]$ is the initial state chosen in the
tracking algorithms.
\end{lyxDefQED}

Correspondingly, we have the following notions associated with the
continuous-time dynamic system. The \emph{equilibrium} of the dynamic
system is defined as the points $\{\widetilde{\bm{\Phi}}_{*}^{c}\in\prod_{b}\mbox{Grass}(m_{b},N_{t})\}$
that satisfy $d\widetilde{\bm{\Phi}}^{c}=0$. An equilibrium $\widetilde{\bm{\Phi}}_{*}^{c}$
is locally \emph{stable} if there exists a neighborhood $\mathcal{N}(\widetilde{\bm{\Phi}}_{*}^{c})\subseteq\prod_{b}\mbox{Grass}(m_{b},N_{t})$,
such that for all $\widetilde{\bm{\Phi}}^{c}(0)\in\mathcal{N}(\widetilde{\bm{\Phi}}_{*}^{c})$,
$\widetilde{\bm{\Phi}}^{c}(t)\to\widetilde{\bm{\Phi}}_{*}^{c}$ as
$t\to\infty$ almost surely (a.s.). In addition, the dynamic system
is called \emph{asymptotically stable} if there exists a unique stable
equilibrium and for all $\widetilde{\bm{\Phi}}^{c}(0)$, we have $\widetilde{\bm{\Phi}}^{c}(t)\to\widetilde{\bm{\Phi}}_{*}^{c}$
as $t\to\infty$ a.s.

Intuitively, scaling the step size $\gamma$ and time slot duration
$\tau$, the discrete-time iteration $\widetilde{\bm{\Phi}}[n]$ degenerates
to its continuous-time counterpart asymptotically as $\tau\to0$.
Such asymptotic equivalence can be established using the stochastic
approximation framework in \cite{Kushner2003vn,Chen13-two}. Specifically,
we summarize the connection in the following (analogue of \cite[Theorem 2]{Chen13-two}).
\begin{lyxLemQED}
\label{lem:Connection-discrete-continuous} \emph{(Connection between
the Algorithm Iteration and the Continuous-time Dynamic System)} Suppose
under each static parameter $\mathcal{Q}$, the dynamic system $\widetilde{\bm{\Phi}}^{c}(t)$
is stable. In addition, assume the channel variation speed $\|d\mathcal{Q}/dt\|$
is bounded above. Then for $\tau,\gamma\to0$ with $\overline{\gamma}=\gamma/\tau\gg1$,
the iterate $\widetilde{\bm{\Phi}}[n]$ converges to the state trajectory
$\widetilde{\bm{\Phi}}^{c}(t)$ of the continuous-time dynamic system
(\ref{eq:continuous-time-dynamic-system}), i.e., for any $\epsilon>0$,
\[
\lim_{\tau,\gamma\to0}\lim\sup_{n\to\infty}\mbox{Pr}\left\{ \|\widetilde{\bm{\Phi}}[n]-\widetilde{\bm{\Phi}}^{c}(t_{n})\|>\epsilon\right\} =0
\]
where $t_{n}=nT_{s}\tau$.
\end{lyxLemQED}

From the above equivalent connection, proving Lemma \ref{lem:unique-stable-equilibrium}
for the iteration (\ref{eq:compensation-algorithm}) under static
channels is equivalent to showing that there is only one stable equilibrium
for the VDS $\widetilde{\bm{\Phi}}^{c}(t)$. This is shown as follows.

From the definition of equilibrium point, we have $\nabla\mathcal{I}(\widetilde{\bm{\Phi}}_{*};\mathcal{Q}(t))=\mathbf{0}$
(see equations (\ref{eq:optimality-cond})). Since $\bm{\Phi}^{[b]}$
and $\bm{\Phi}^{[l]}$ do not couple in $\mathcal{I}(\widetilde{\bm{\Phi}};\mathcal{Q})$,
we have $\nabla_{\bm{\Phi}_{*}^{[b]}}\mathcal{I}(\widetilde{\bm{\Phi}}_{*})=\mathbf{0}$,
which leads to 
\[
\mathbf{Q}^{[b]}\bm{\Phi}_{*}^{[b]}=\bm{\Phi}_{*}^{[b]}\bm{\Phi}_{*}^{[b]\dagger}\mathbf{Q}^{[b]}\bm{\Phi}_{*}^{[b]},\qquad\forall b
\]
due to the gradient equation (\ref{eq:gradient}) and the fact that
$\bm{\Phi}^{[b]\dagger}\bm{\Phi}^{[b]}=\mathbf{I}$ for Algorithm
1. As a result, $\mbox{span}(\bm{\Phi}_{*}^{[b]})$ must span the
eigen subspace of $\mathbf{Q}^{[b]}$. That means the columns of $\hat{\bm{\Phi}}_{*}^{[b]}=\bm{\Phi}_{*}^{[b]}\mathbf{M}$,
where $\mathbf{M}$ is an appropriately chosen unitary matrix, are
the eigenvectors of $\mathbf{Q}^{[b]}$. 

Using the Lyapunov stability analysis techniques \cite{Khalil1996},
consider a Lyapunov function $V(\widetilde{\bm{\Phi}}^{e})=\frac{1}{2}\mbox{tr}\left[\widetilde{\bm{\Phi}}^{e\dagger}\widetilde{\bm{\Phi}}^{e}\right]$.
We write $\widetilde{\bm{\Phi}}^{c}=(\bm{\Phi}_{c}^{[1]},\dots,\bm{\Phi}_{c}^{[b]},\dots,\bm{\Phi}_{c}^{[G]})$
and $\widetilde{\bm{\Phi}}^{e}=(\bm{\Phi}_{e}^{[1]},\dots,\bm{\Phi}_{e}^{[b]},\dots,\bm{\Phi}_{e}^{[G]})$
for each component $b$, where $\bm{\Phi}_{e}^{[b]}=\bm{\Phi}_{c}^{[b]}-\bm{\Phi}_{*}^{[b]}$.
We have 
\begin{align}
\dot{V}(\widetilde{\bm{\Phi}}^{e}) & =\mbox{tr}\left\{ \mbox{Re}\left[\widetilde{\bm{\Phi}}^{e\dagger}d\widetilde{\bm{\Phi}}^{e}\right]\right\} \label{eq:V-dot}\\
 & \stackrel{(a)}{=}-\mbox{tr}\left\{ \mbox{Re}\left[\widetilde{\bm{\Phi}}^{e\dagger}F(\widetilde{\bm{\Phi}}^{c};\mathcal{Q}(t))\right]\right\} \nonumber \\
 & \stackrel{(b)}{=}-\sum_{b=1}^{G}\int_{0}^{1}(1-\mu)\mbox{tr}\left[\bm{\Phi}_{e}^{[b]\dagger}\mbox{Hess}_{(\bm{\Phi}_{*}^{[b]}+\mu\bm{\Phi}_{e}^{[b]})}(\bm{\Phi}_{e}^{[b]})\right]d\mu\nonumber 
\end{align}
where equation $\stackrel{(a)}{=}$ is due to $d\widetilde{\bm{\Phi}}_{*}^{c}=\mathbf{0}$
under static channels and equation $\stackrel{(b)}{=}$ is obtained
from the multi-dimensional Taylor's expansion of $F(\widetilde{\bm{\Phi}}^{c};\mathcal{Q}(t))$.

To study the term in $\stackrel{(b)}{=}$, without loss of generality,
we consider the columns of $\bm{\Phi}_{*}^{[b]}$ are the eigenvectors
of $\mathbf{Q}^{[b]}$. From (\ref{eq:Hessian-long}), we have 
\begin{equation}
\mbox{Hess}_{\bm{\Phi}_{*}^{[b]}}(\bm{\Phi}_{e}^{[b]})=\left(\mathbf{I}-\bm{\Phi}_{*}^{[b]}\bm{\Phi}_{*}^{[b]\dagger}\right)\left[\mathbf{Q}^{[b]}(\bm{\Phi}_{e}^{[b]})-(\bm{\Phi}_{e}^{[b]})\bm{\Psi}^{[b]}\right]\label{eq:app-Hessian-matrix}
\end{equation}
where $\bm{\Psi}^{[b]}$ is a diagonal matrix with diagonal elements
$\lambda_{i}^{[b]}$ being the eigenvalues of $\mathbf{Q}^{[b]}$
and $\mathbf{I}-\bm{\Phi}_{*}^{[b]}\bm{\Phi}_{*}^{[b]\dagger}$ is
a projection matrix onto the null space of eigen subspace $\mbox{span}(\bm{\Phi}_{*}^{[b]})$.
Therefore, the $i$-th column of the Hessian (\ref{eq:app-Hessian-matrix})
is 
\begin{equation}
\mbox{Hess}_{\bm{\Phi}_{*}^{[b]}}(\bm{\Phi}_{e}^{[b],i})=\left(\mathbf{I}-\bm{\Phi}^{[b]}\bm{\Phi}^{[b]\dagger}\right)\left(\mathbf{Q}^{[b]}-\lambda_{i}\mathbf{I}\right)(\bm{\Phi}_{e}^{[b],i})\label{eq:app-Hessian-optimal-point}
\end{equation}
where $\bm{\Phi}_{e}^{[b],i}$ denotes the $i$-th column of $\bm{\Phi}_{e}^{[b]}$,
and the eigenvalues of the matrix $\left(\mathbf{I}-\bm{\Phi}^{[b]}\bm{\Phi}^{[b]\dagger}\right)\left(\mathbf{Q}^{[b]}-\lambda_{i}\mathbf{I}\right)$
are either $0$ (due to the eigen subspace projection) or $\lambda_{j}^{[b]}-\lambda_{i}^{[b]}$,
where $\lambda_{j}^{[b]}$ are the eigenvalues of $\mathbf{Q}^{[b]}$
corresponding to the eigenvectors of the null space of $\bm{\Phi}_{*}^{[b]}$.

Therefore, the Lyapunov drift $\dot{V}(\widetilde{\bm{\Phi}}^{e})$
in (\ref{eq:V-dot}) is negative semi-definite if and only if $\lambda_{j}^{[b]}-\lambda_{i}^{[b]}\geq0$
for all $j$, which means $\bm{\Phi}_{*}^{[b]}$ spans the minimum
eigen subspace of $\mathbf{Q}^{[b]}$ for each $b$. In other words,
under such choice of $\bm{\Phi}_{*}^{[b]}$, the Lyapunov function
$V(\widetilde{\bm{\Phi}}^{e})$ is always decreasing, unless it reaches
the point $\widetilde{\bm{\Phi}}^{e}=\mathbf{0}$. As a result, $\widetilde{\bm{\Phi}}_{*}$
gives a stable equilibrium. 

Note that, due to the distinct eigenvalue condition for $\mathbf{Q}^{[b]}$,
the minimum eigen subspace for each $\mathbf{Q}^{[b]}$ is unique.
Hence, such stable equilibrium is unique.



%

As a result, under static channel covariance, the VDS has only one
stable equilibrium and it would converge to the unique stable equilibrium
w.p.1. This is because when the state trajectory of the VDS passes
over the neighborhood of an unstable equilibrium, there is a high
probability for it to be expelled away from the unstable equilibrium
and it can only be attracted by the unique stable equilibrium.

\subsection{Proof of Theorem \ref{thm:global-convergence}}

The global attraction property of Algorithm 1 has been established
by Lemma \ref{lem:unique-stable-equilibrium} using an equivalent
continuous-time trajectory. Therefore, Algorithm 1 is guaranteed to
converge to the neighborhood of the global unique optimal point $\widetilde{\bm{\Phi}}_{*}$
even for non-decreasing step size $\gamma>0$. Moreover, since $\mathcal{I}(\widetilde{\bm{\Phi}})$
is Lipschitz continuous in $\widetilde{\bm{\Phi}}$, using \cite[Proposition 3.2.1]{Bertsekas:1989fk},
for sufficiently small step size $0<\gamma<\gamma_{0}$, the gradient
algorithm converges to the unique global optimal point under static
channels.

\section{Proof of Theorem \ref{thm:convergence-compensation}}

\label{app:pf-thm-convergence-compensation}

Applying Hessian operator at the point $\widetilde{\bm{\Phi}}^{c}$
to the above compensation algorithm flow (\ref{eq:compensation-alg-flow}),
we have $\mbox{Hess}_{\widetilde{\bm{\Phi}}^{c}}(d\widetilde{\bm{\Phi}}^{c})=\mbox{Hess}_{\widetilde{\bm{\Phi}}^{c}}(F(\widetilde{\bm{\Phi}}^{c};\mathcal{Q}(t))dt+\mbox{Hess}_{\widetilde{\bm{\Phi}}^{c}}(\widehat{d\widetilde{\bm{\Phi}}_{*}}).$
From the optimality condition (\ref{eq:continuous-time-compensation-flow}),
we have 
\begin{equation}
\begin{array}{l}
\mbox{Hess}_{\widetilde{\bm{\Phi}}^{c}}(d\widetilde{\bm{\Phi}}^{c})+F(\widetilde{\bm{\Phi}}^{c};d\mathcal{Q})\\
\quad=\quad\mbox{Hess}_{\widetilde{\bm{\Phi}}^{c}}(F(\widetilde{\bm{\Phi}}^{c};\mathcal{Q}(t))dt+\mbox{Hess}_{\widetilde{\bm{\Phi}}^{c}}(\widehat{d\widetilde{\bm{\Phi}}_{*}})+F(\widetilde{\bm{\Phi}}^{c};d\mathcal{Q})\\
\quad=\quad\mbox{Hess}_{\widetilde{\bm{\Phi}}^{c}}(F(\widetilde{\bm{\Phi}}^{c};\mathcal{Q}(t))dt.
\end{array}\label{eq:app-hessian-property}
\end{equation}

As an intermediate result, we want to show that the gradient mapping
$F(\widetilde{\bm{\Phi}}^{c};\mathcal{Q}(t))$ converges to zero,
which is a necessary condition for $\widetilde{\bm{\Phi}}^{c}(t)$
converging to $\widetilde{\bm{\Phi}}_{*}(t)$. In the following, we
construct a Lyapunov function $V(\widetilde{\bm{\Phi}}^{c};t)=\frac{1}{2}\mbox{tr}\left[F(\widetilde{\bm{\Phi}}^{c};\mathcal{Q}(t))^{\dagger}F(\widetilde{\bm{\Phi}}^{c};\mathcal{Q}(t))\right]$
and check the property of the \emph{Lyapunov drift} $\dot{V}$.

Using the chain rule gives $dF(\widetilde{\bm{\Phi}}^{c}(t);\mathcal{Q}(t))=\mbox{Hess}_{\widetilde{\bm{\Phi}}^{c}}(d\widetilde{\bm{\Phi}}^{c})+F(\widetilde{\bm{\Phi}}^{c};d\mathcal{Q})$,
we have 
\begin{align*}
\dot{V}(\widetilde{\bm{\Phi}}^{c};t) & =\mbox{tr}\left\{ \mbox{Re}\left[F(\widetilde{\bm{\Phi}}^{c};\mathcal{Q}(t))^{\dagger}dF(\widetilde{\bm{\Phi}}^{c}(t);\mathcal{Q}(t))\right]\right\} \\
 & =\mbox{tr}\left\{ \mbox{Re}\left[F(\widetilde{\bm{\Phi}}^{c};\mathcal{Q}(t))^{\dagger}\left(\mbox{Hess}_{\widetilde{\bm{\Phi}}^{c}}\widetilde{\bm{\Phi}}^{e}+F(\widetilde{\bm{\Phi}}^{c},d\mathcal{Q})\right)\right]\right\} \\
 & =\mbox{tr}\left\{ \mbox{Re}\left[F(\widetilde{\bm{\Phi}}^{c};\mathcal{Q}(t))^{\dagger}\mbox{Hess}_{\widetilde{\bm{\Phi}}^{c}}(F(\widetilde{\bm{\Phi}}^{c};\mathcal{Q}(t))\right]\right\} 
\end{align*}
where the last equality is from (\ref{eq:app-hessian-property}).

Note that the function $F(\centerdot)$ defined in (\ref{eq:optimality-cond})
is a projected gradient of the quadratic function $\mathcal{I}(\centerdot)$
defined in (\ref{prob:objective-Grassmann}). Therefore, $F(\centerdot)$
is Lipschitz continuous. Thus, when $\tilde{\bm{\Phi}}^{c}$ is close
enough to $\widetilde{\bm{\Phi}}_{*}$ (i.e., $\|\widetilde{\bm{\Phi}}^{e}\|$
is small), $\|F(\widetilde{\bm{\Phi}}^{c};\mathcal{Q}(t))\|$ is close
to $\|F(\widetilde{\bm{\Phi}}_{*};\mathcal{Q}(t))\|$, which equals
to 0. 

From the analysis of the Hessian (\ref{eq:app-Hessian-optimal-point})
in Appendix \ref{app:VDS}, $\eta^{\dagger}\mbox{Hess}_{\bm{\Phi}_{*}^{[b]}}(\eta)$
is a negative semi-definite quadratic form, provided that $\lambda_{m+1}-\lambda_{m}>0$.
Therefore, $\mbox{tr}\big\{\mbox{Re}\big[F(\widetilde{\bm{\Phi}}^{c};\mathcal{Q}(t))^{\dagger}\mbox{Hess}_{\widetilde{\bm{\Phi}}^{c}}(F(\widetilde{\bm{\Phi}}^{c};\mathcal{Q}(t))\big]\big\}$
is negative definite as long as $\tilde{\bm{\Phi}}^{c}$ is sufficiently
close to $\tilde{\bm{\Phi}}_{*}$. Mathematically, there exists $\epsilon_{1}>0$,
such that for all $\widetilde{\bm{\Phi}}$ in the $\epsilon_{1}$-neighborhood
of $\widetilde{\bm{\Phi}}_{*}$, $\widetilde{\bm{\Phi}}\in\mathcal{N}_{\epsilon_{1}}(\widetilde{\bm{\Phi}}_{*}(t))=\{\widetilde{\bm{\Phi}}:\|\widetilde{\bm{\Phi}}-\widetilde{\bm{\Phi}}_{*}(t)\|_{F}<\epsilon_{1}\}$,
we have (i) $\dot{V}(\widetilde{\bm{\Phi}}^{c};t)<0$, and (ii) an
inverse mapping $F^{-1}:T_{X}\mbox{Grass}(p,N_{t})\mapsto\mbox{Grass}(p,n_{t})$
satisfying for all $t$, under some $k_{1}<\infty$, 
\begin{eqnarray*}
\|\widetilde{\bm{\Phi}}-\widetilde{\bm{\Phi}}_{*}(t)\|_{F} & = & \|F^{-1}(F(\widetilde{\bm{\Phi}}))-F^{-1}(F(\widetilde{\bm{\Phi}}_{*}(t)))\|_{F}\\
 & \leq & k_{1}\|F(\widetilde{\bm{\Phi}})-F(\widetilde{\bm{\Phi}}_{*}(t))\|_{F}.
\end{eqnarray*}

The property $\dot{V}<0$ implies that $\|F(\widetilde{\bm{\Phi}}^{c};\mathcal{Q}(t))\|$
always decreases whenever $\|F\|\neq0$, from the Lyapunov stability
theory \cite{Khalil1996}. Moreover, since the largest eigenvalue
$\lambda_{\max}$ is bounded, there exists $k_{2}<\infty$, such that
$\|F(\widetilde{\bm{\Phi}})-F(\widetilde{\bm{\Phi}}_{*}(t))\|_{F}<k_{2}\|\widetilde{\bm{\Phi}}-\widetilde{\bm{\Phi}}_{*}(t)\|_{F}$
for all $t$.

As a result, choosing $\delta<\frac{\epsilon_{1}}{k_{1}k_{2}}$ and
$\|\widetilde{\bm{\Phi}}^{c}(0)-\widetilde{\bm{\Phi}}_{*}(0)\|_{F}<\delta$,
we must have $\|\widetilde{\bm{\Phi}}^{c}(t)-\widetilde{\bm{\Phi}}_{*}(t)\|_{F}<k_{1}k_{2}\|\widetilde{\bm{\Phi}}^{c}(0)-\widetilde{\bm{\Phi}}_{*}(0)\|_{F}<k_{1}k_{2}\delta$
and $\widetilde{\bm{\Phi}}^{c}(t)\in\mathcal{N}_{\epsilon_{1}}(\widetilde{\bm{\Phi}}_{*}(t))$
for all $t$. Therefore, $\|\widetilde{\bm{\Phi}}^{c}(t)-\widetilde{\bm{\Phi}}_{*}(t)\|_{F}<k_{1}\|F(\widetilde{\bm{\Phi}}^{c};\mathcal{Q}(t))\|_{F}\to0$.

\bibliographystyle{IEEEtran}
\bibliography{/Users/Allen/Dropbox/Draft/My_reference}

\begin{biography}
[{\includegraphics[width=1in,height=1.25in,clip,keepaspectratio]{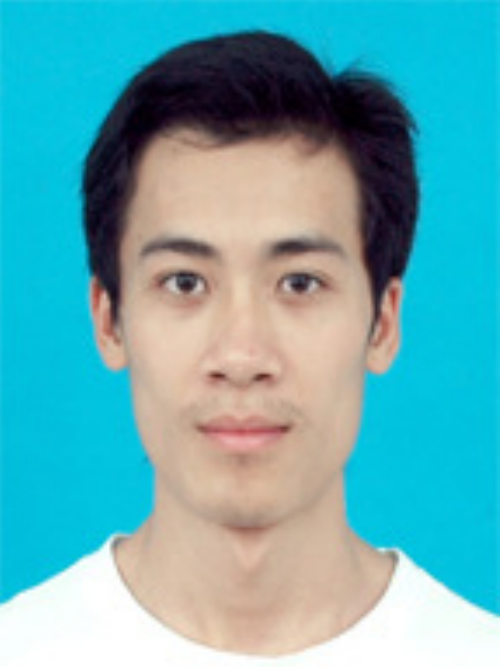}}]{Junting
Chen (S'11)} received the B.Sc. degree in electronic science and
technology from Nanjing University, Nanjing, China, in 2009. He is
with the Department of Electronic and Computer Engineering, The Hong
Kong University of Science and Technology (HKUST), Hong Kong, where
he is now a Ph.D. candidate.

Since February 2014, he has been in the Laboratory for Information
and Decision Systems (LIDS) at Massachusetts Institute of Technology
(MIT) as a visiting student. His research interests include beamformer
design in massive MIMO systems, resource allocations and cross-layer
optimizations in wireless communication networks, and algorithm design
and analysis under time-varying channels.
\end{biography}
\begin{biography}
[{\includegraphics[width=1in,height=1.25in,clip,keepaspectratio]{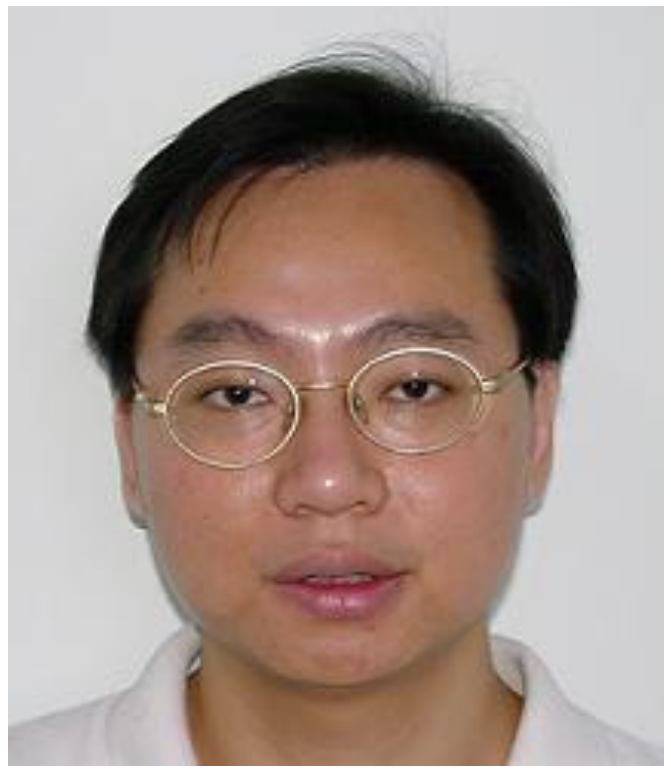}}]{Vincent
K. N. Lau (SM'04)} received the B.Eng. (Distinction 1st Hons.) from
the University of Hong Kong in 1992 and the Ph.D. degree from Cambridge
University, Cambridge, U.K., in 1997.

He was with HK Telecom (PCCW) as a System Engineer from 1992 to 1995,
and with Bell Labs - Lucent Technologies as a member of Technical
Staff during 1997-2003. He then joined the Department of ECE, HKUST,
and is currently a Professor. His current research interests include
the robust and delay-sensitive cross-layer scheduling of MIMO/OFDM
wireless systems, cooperative and cognitive communications, dynamic
spectrum access, as well as stochastic approximation and Markov decision
process.\end{biography}

\end{document}